\documentclass[10pt,journal,compsoc]{IEEEtran}

\usepackage{amsmath,amsfonts}
\usepackage{algorithmic}
\usepackage{algorithm}
\usepackage{array}
\usepackage[caption=false,font=normalsize,labelfont=sf,textfont=sf]{subfig}
\usepackage{textcomp}
\usepackage{stfloats}
\usepackage{url}
\usepackage{verbatim}
\usepackage{graphicx}
\usepackage{pdfpages}

\usepackage{makecell}
\usepackage{amssymb}
\usepackage{booktabs}
\usepackage{multirow}
\usepackage{xspace}
\usepackage{bm}
\usepackage{epsfig}
\usepackage{color}
\usepackage{comment}
\usepackage{arydshln}
\usepackage{dcolumn}
\usepackage{eqparbox}
\usepackage{subcaption}
\usepackage{pifont}
\usepackage{enumitem}
\ifCLASSOPTIONcompsoc
  \usepackage[nocompress]{cite}
\else
  \usepackage{cite}
\fi

\ifCLASSINFOpdf

\else

\fi

\begin{document}

\title{Towards Social AI: \\A Survey on Understanding Social Interactions}

\author{Sangmin Lee, Minzhi Li, Bolin Lai, Wenqi Jia, Fiona Ryan, Xu Cao, Ozgur Kara, Bikram Boote, \\ Weiyan Shi, Diyi Yang, and James M. Rehg

\IEEEcompsocitemizethanks{


\IEEEcompsocthanksitem Sangmin Lee, Wenqi Jia, Xu Cao, Ozgur Kara, Bikram Boote, and James M. Rehg are with the University of Illinois Urbana-Champaign, Urbana, IL 61801, USA (E-mail: sangminl@illinois.edu, wenqij5@illinois.edu, xucao2@illinois.edu, ozgurk2@illinois.edu, boote@illinois.edu, jrehg@illinois.edu)

\IEEEcompsocthanksitem Minzhi Li is with National University of Singapore, 117417, Singapore (E-mail: li.minzhi@u.nus.edu)

\IEEEcompsocthanksitem Bolin Lai, and Fiona Ryan are with the Georgia Institute of Technology, Atlanta, GA 30308, USA (E-mail: bolin.lai@gatech.edu, fkryan@gatech.edu)

\IEEEcompsocthanksitem Weiyan Shi and Diyi Yang are with Stanford, Stanford, CA 94305, USA (E-mail: weiyans@stanford.edu, diyiy@cs.stanford.edu)
}
}

\markboth{Journal of \LaTeX\ Class Files,~Vol.~14, No.~8, August~2015}%
{Shell \MakeLowercase{\textit{et al.}}: Bare Demo of IEEEtran.cls for Computer Society Journals}

\IEEEtitleabstractindextext{%
\begin{abstract}
Social interactions form the foundation of human societies. Artificial intelligence has made significant progress in certain areas, but enabling machines to seamlessly understand social interactions remains an open challenge. It is important to address this gap by endowing machines with social capabilities. We identify three key capabilities needed for effective social understanding: 1) understanding multimodal social cues, 2) understanding multi-party dynamics, and 3) understanding beliefs. Building upon these foundations, we classify and review existing machine learning works on social understanding from the perspectives of verbal, non-verbal, and multimodal social cues. The verbal branch focuses on understanding linguistic signals such as speaker intent, dialogue sentiment, and commonsense reasoning. The non-verbal branch addresses techniques for perceiving social meaning from visual behaviors such as body gestures, gaze patterns, and facial expressions. The multimodal branch covers approaches that integrate verbal and non-verbal multimodal cues to holistically interpret social interactions such as recognizing emotions, conversational dynamics, and social situations. By reviewing the scope and limitations of current approaches and benchmarks, we aim to clarify the development trajectory and illuminate the path towards more comprehensive intelligence for social understanding. We hope this survey will spur further research interest and insights into this area.
\end{abstract}

\begin{IEEEkeywords}
Social interaction, verbal, non-verbal, multimodal, multi-party, belief.
\end{IEEEkeywords}}

\maketitle

\IEEEdisplaynontitleabstractindextext

\IEEEpeerreviewmaketitle

\IEEEraisesectionheading{\section{Introduction}\label{sec:introduction}}
\IEEEPARstart{S}{ocial} interactions are an integral and fundamental part of human societies. In our daily lives, we are constantly immersed in social interactions - engaging in conversations and collaborating with others. Humans have a remarkable ability to interpret subtle social cues and navigate complex interpersonal dynamics \cite{frith2012mechanisms}. This allows us to understand intentions, emotions, and situational contexts, and consequently build relationships with others effectively.
 
Social intelligence of machines to understand social interactions can be the foundation for developing various AI applications. Embodied agents \cite{wu2023tidybot} or virtual agents \cite{park2023generative} require social understanding capabilities to effectively interact and collaborate with humans. More broadly, capabilities for social understanding can enhance diverse types of AI agents, from chatbots \cite{shum2018eliza} to recommendation systems \cite{chen2019serendipity}, to better understand user needs and build rapport with humans. By integrating social understanding into systems, we can create more effective and socially adept AI applications across diverse domains.
 
Artificial intelligence has made significant progress in recent years, achieving human-level performance or beyond on specialized tasks in certain areas of natural language processing and computer vision~\cite{achiam2023gpt,team2023gemini,Kirillov_2023_ICCV,rombach2022high}. However, enabling machines to engage naturally alongside humans in social contexts remains an open challenge \cite{yang2024socially}. It is important to address this gap by endowing machines with social capabilities to understand human interactions seamlessly. 

To identify the key capabilities required for machines to understand social interactions, we can consider the example of social deduction games \cite{lai2023werewolf, chittaranjan2010you}, where people take on specific roles and try to reveal the hidden roles of their opponents. These games require players to engage in communication, deception, inference, and collaboration, encompassing rich social interactions. For instance, player A may think that players B and C are allies working together against the enemy based on their willingness to trust each other. On the other hand, player D seems deceptive and evasive, leading A to suspect they are an enemy pretending to be an ally. To reach this judgment, player A relies on interpreting social interactions over multiple discussion rounds. This example illustrates three key capabilities needed for effective social understanding:

\begin{figure*}[t]
	\begin{minipage}[b]{1.0\linewidth}
		\centering
		\centerline{\includegraphics[width=18.5cm]{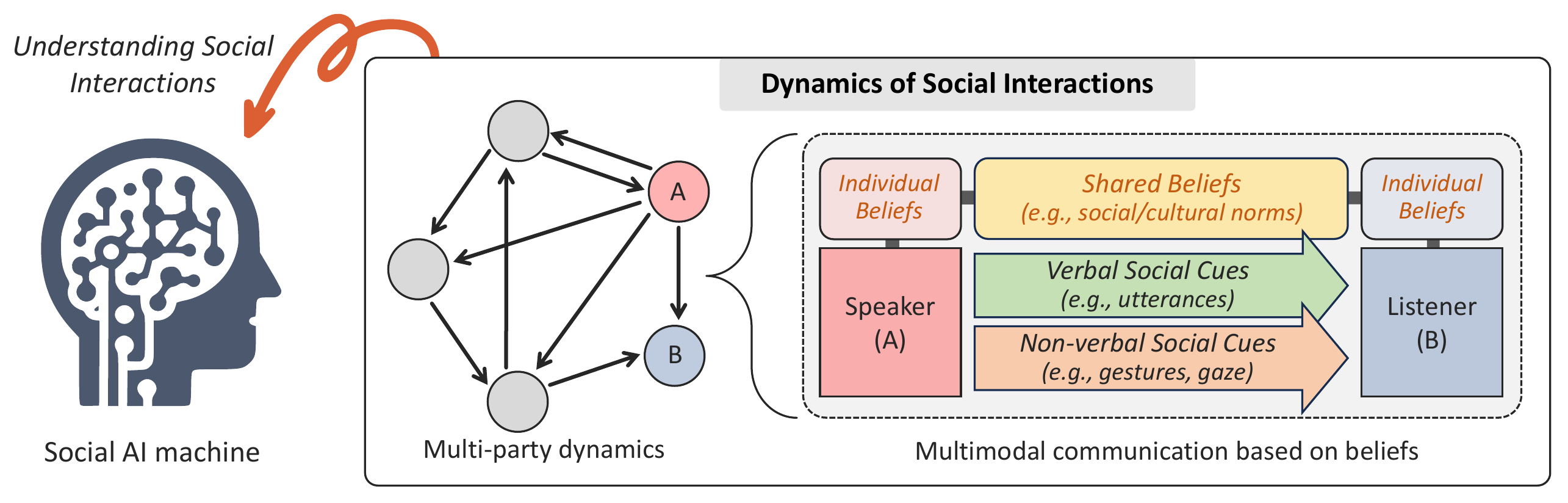}}
	\end{minipage}
	\vspace{-0.6cm}
	\captionof{figure}{Dynamics of social interactions, illustrating three key capabilities: multimodal understanding, multi-party modeling, and belief awareness. Machines need to be equipped with these capabilities to effectively interpret social meaning regarding intentions, emotions, and situational contexts in social interactions.}
	\label{figure_dynamics}
	\vspace{-0.30cm}
\end{figure*}

\begin{enumerate}
\item  \textbf{Understanding multimodal social cues:} This involves interpreting verbal utterances along with non-verbal behaviors such as facial expressions and body language comprehensively. While verbal communication conveys rich semantic meaning, non-verbal cues often clarify subtle social nuances like detailed sentiments and emotions. They can also indicate the target of an utterance through gestures and gaze. Therefore, holistically recognizing these multimodal clues provides a basis for understanding social interactions.

\item  \textbf{Understanding multi-party dynamics:} Social interactions often involve multiple participants. Analyzing group conversations requires aggregating individual social behaviors and modeling the interpersonal connections between people. This entails multifaceted aspects of analyzing dynamics such as tracking who is speaking to whom, who is interrupting whom, and who is being positive to whom. Based on the interpretation of such relationships, we can effectively understand social interactions among multiple people.

\item  \textbf{Understanding beliefs:} Humans have complex mental models comprised of individual and shared beliefs, which influence social interactions. Individual beliefs are relevant to individual personalities and dispositions while shared beliefs associated with what people share such as social norms, cultural contexts, social relationships, and game rules. Being aware of these beliefs in social interactions enables appropriate social reasoning. 
\end{enumerate}

As a result, the roadmap for ideal social AI involves building machines with capabilities across those three fronts: multimodal understanding, multi-party modeling, and belief awareness. Figure \ref{figure_dynamics} shows the dynamics of social interactions including these capabilities. Through these, machines can approach human social intelligence for seamlessly interpreting social interactions. While progress has been made in narrow applications, advanced social intelligence requires integrating breakthroughs across those areas. 

\begin{figure*}[t]
	\begin{minipage}[b]{1.0\linewidth}
		\centering
		\centerline{\includegraphics[width=18.5cm]{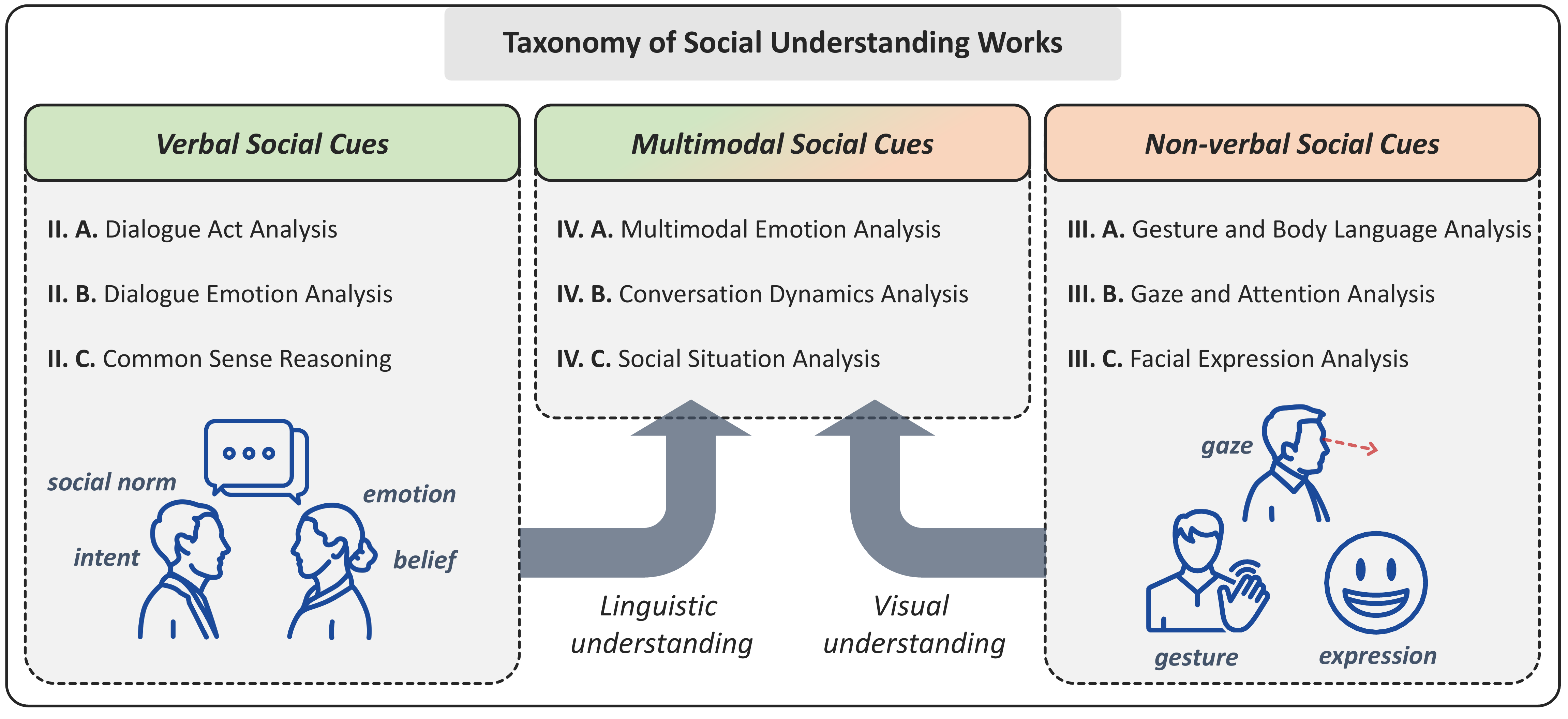}}
	\end{minipage}
	\vspace{-0.6cm}
	\caption{Taxonomy of existing research on social understanding organized according to social cue types, such as verbal and non-verbal cues. The taxonomy covers studies on linguistic understanding from dialogues, visual perception of non-verbal behaviors, and joint understanding of verbal and non-verbal cues.}
	\label{figure_taxonomy}
	\vspace{-0.30cm}
\end{figure*}

We classify existing works related to understanding social interactions from the perspective of social cue types (\textit{i.e.}, verbal and non-verbal cues). This classification is associated with how to extract social meaning from each social cue, which can provide a foundation for holistically understanding social interactions. Figure \ref{figure_taxonomy} presents a taxonomy of existing research on social understanding, organized according to the types of social cues. The taxonomy includes three main branches: verbal cues, non-verbal cues, and multimodal cues. Under the verbal cues branch, key research areas include dialogue act analysis for recognizing intent, dialogue emotion analysis, and commonsense reasoning to interpret socially common knowledge in language. The non-verbal cues branch encompasses research on recognizing non-verbal behaviors including gestures, gaze, and facial expressions. Finally, the multimodal branch covers works that integrate multimodal signals to more comprehensively infer emotions, conversation dynamics, and social situations.

The components listed under each category in Figure \ref{figure_taxonomy} are not meant to be isolated research areas. They represent key building blocks and capabilities that need to be connected and integrated to enable truly holistic social understanding. For example, dialogue act analysis and emotion recognition from facial expressions could be combined to better interpret the intent and sentiment behind an utterance. Similarly, gesture recognition and gaze analysis could be jointly leveraged with spoken language to recognize who an utterance is directed at in a multi-party conversation. Therefore, while organizing the existing research, the taxonomy also points towards the need for further integration across areas. In the course of our survey, we will touch upon this need for integration of multimodal social cues, as well as discuss how multi-party dynamics and belief modeling intersect with the social understanding landscape.

This is the first survey that addresses a comprehensive overview of machine learning studies for social understanding, covering both verbal and non-verbal approaches. While there have been surveys addressing specific aspects of social understanding such as gaze analysis \cite{ghosh2023automatic}, dialogue systems \cite{chen2017survey}, and facial expression recognition \cite{ben2021video}, these works focus on narrow subdomains without considering the broader social context. There also have been position papers on social AI regarding data infrastructure \cite{li2024social} and social AI agents \cite{mathur2024advancing}. While these papers provide valuable high-level discussions of relevant works, they do not focus on offering a detailed, comprehensive survey of existing literatures. In contrast, our paper offers a holistic survey of the social understanding landscape, encompassing techniques across multiple modalities and identifying key capabilities crucial for advancing socially intelligent AI. By organizing the wide range of technical works around these core capabilities, we aim to provide a unified perspective and illuminate promising research directions. The scope of our survey covers the problem of social understanding rather than the creation of agents, as we argue that social understanding is a fundamental prerequisite for socially intelligent agents. 

We structure our paper as follows. In Sections 2-4, we provide an overview of recent advancements related to understanding verbal, non-verbal, and multimodal social cues, respectively. Section 5 discusses open challenges and promising future research directions based on the suggested key capabilities. Finally, we summarize the key points of this survey and offer concluding remarks in Section 6. By reviewing existing techniques and limitations, we aim to clarify the development trajectory and road ahead for social understanding research.

\section{Understanding Verbal Cues}

\subsection{Dialogue Act Analysis}
\subsubsection{\textbf{Social Background}} A dialogue act is ``a combination of a communicative function and a semantic content" \cite{bunt2000dialogue}. It refers to the intended action and communicative purpose of an utterance. For instance, when asking for help to open a door, various utterances can fulfill the same dialogue act. This analysis is crucial for social AI systems in natural language understanding, intent detection, and contextual awareness.

There are different planes and dimensions for communicative functions in the utterances \cite{stolcke2000dialogue}. Some dimensions stem from speakers' mental states \cite{searle1976classification, vanderveken1990meaning} and taxonomize dialogue acts into \textit{Representatives} (assertion, conclusion, etc.), \textit{Directives} (request, question, suggestion, etc.), \textit{Commissives} (promise, threat, offer, etc.), \textit{Expressives}
(thanks, apologize, welcome, congratulation, etc.) and \textit{Declarations} (christening, firing from employment, excommunication, declaration of war, etc.). Different dialogue acts can also be categorized based on conversation management \cite{sacks1978simplest, clark1989contributing}, with categorization such as \textit{take turn}, \textit{maintain turn}, \textit{give turn}, and \textit{let others maintain the turn}. Moreover, dialogue acts can be defined with rhetorical roles and relations \cite{grosz1986attention, mann1988rhetorical} (\emph{e.g.}, \textit{purpose}, \textit{condition}, \textit{contrast} etc.) or the notion of ``face" \cite{brown1987politeness} (\textit{face saving} and \textit{face threatening}). In addition, some work recognizes functions of pared utterances and introduce types for such adjacency pairs \cite{levinson1983pragmatics, schegloff1973opening} like \textit{assess $\rightarrow$ agree/disagree}, \textit{question $\rightarrow$ answer}, \textit{thank $\rightarrow$ welcome}, \textit{offer $\rightarrow$ accept/refuse} and so on.


\begin{figure}[t]
	\begin{minipage}[b]{1.0\linewidth}
		\centering
		\centerline{\includegraphics[width=9.0cm]{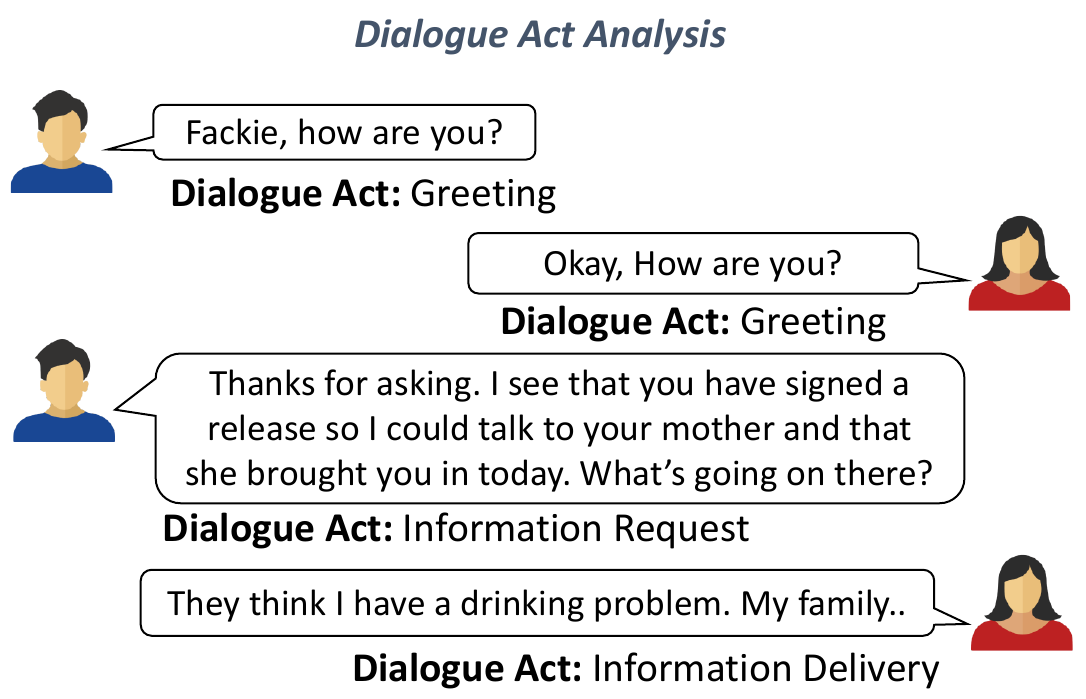}}
	\end{minipage}
	\vspace{-0.6cm}
	\caption{Examples of dialogue act analysis from HOPE \cite{malhotra2022speaker}. Dialogue act analysis involves classifying the speaker's intention or communicative goal behind the utterance. }
	\label{figure_dialogue_act}
	\vspace{-0.30cm}
\end{figure}

\subsubsection{\textbf{Dialogue Act Classification}} Early work treated dialogue act classification as a sequence labeling problem and applied artificial neural networks for dialogue act prediction. Lee and Dernoncourt \cite{lee2016sequential} proposed a model based on CNNs and RNNs. Later work used hierarchical neutral networks for the task. More recent work incorporates context-aware self attention mechanism to the hierarchical neural network to further enhance dialogue act understanding. Raheja and Tetreault \cite{raheja2019dialogue} combined self-attention, hierarchical neural networks and contextual dependencies to develop a new model that leads to performance improvements. Li \emph{et al.} \cite{li2019dual} recognized the association between dialogue acts and subject matters and introduced dual-attention hierarchical RNN to capture such dependency. Chapuis \emph{et al.} \cite{chapuis2020hierarchical} used a hierarchical encoder based on transformer architectures to learn generic representations of dialogues in the pretraining stage. Colombo \emph{et al.} \cite{colombo2020guiding} proposed a sequence-to-sequence model with a hierarchical encoder, guided attention mechanism and beam search method. Moreover, work like \cite{he2021speaker} considered the interactive nature of dialogue by incorporating turn changes among speakers into dialogue act modeling. Malhotra \emph{et al.} \cite{malhotra2022speaker} also introduced speaker- and time-aware contextual learning in their counseling conversation dataset. As such, information about speakers can be captured and it provides more semantic information about the conversations. Furthermore, understanding individuals' beliefs helps with dialogue state tracking. Mrksic \emph{et al.} \cite{mrkvsic2017neural} leveraged this idea with a neural belief tracker to predict goals in dialogues.

On top of model architectures that handle textual data, researchers also came up various approaches for data with multiple modalities. He \emph{et al.} \cite{he2018exploring} leveraged CNNs for audio feature augmentation and RNNs for utterance modeling. However, the way information from different modalities are fused is by concatenation, which is not so effective. Given the challenge of integrating representations of different modalities, Miah \emph{et al.} \cite{miah2023hierarchical} proposed a novel framework that works for raw audio and ASR-generated transcriptions and combines different modalities at a more granular level.

\subsubsection{\textbf{Discussion}} Dialogue act analysis is important for understanding dialogue contexts and the information exchange among different parties. However, it often tends to focus on the local context of the dialogue exchange itself, without fully accounting for the broader social context in which the interaction is embedded. Social factors like power relations and cultural norms can shape dialogue acts in important ways that may be missed. Additionally, much of the research has focused on the verbal aspects of interaction, despite face-to-face conversations involving rich multimodal signals like gaze and gesture. Furthermore, most dialogue act analysis works are based on task-oriented or single-topic interactions between two participants, whereas real-world social interactions often involve multiple participants discussing various topics and goals, adding complexity. Analyzing dialogues with multiple participants further complicates the task as these often lack clear turn-taking and involve overlapping speech, making it difficult to identify dialogue acts well.


\begin{figure}[t]
	\begin{minipage}[b]{1.0\linewidth}
		\centering
		\centerline{\includegraphics[width=9.0cm]{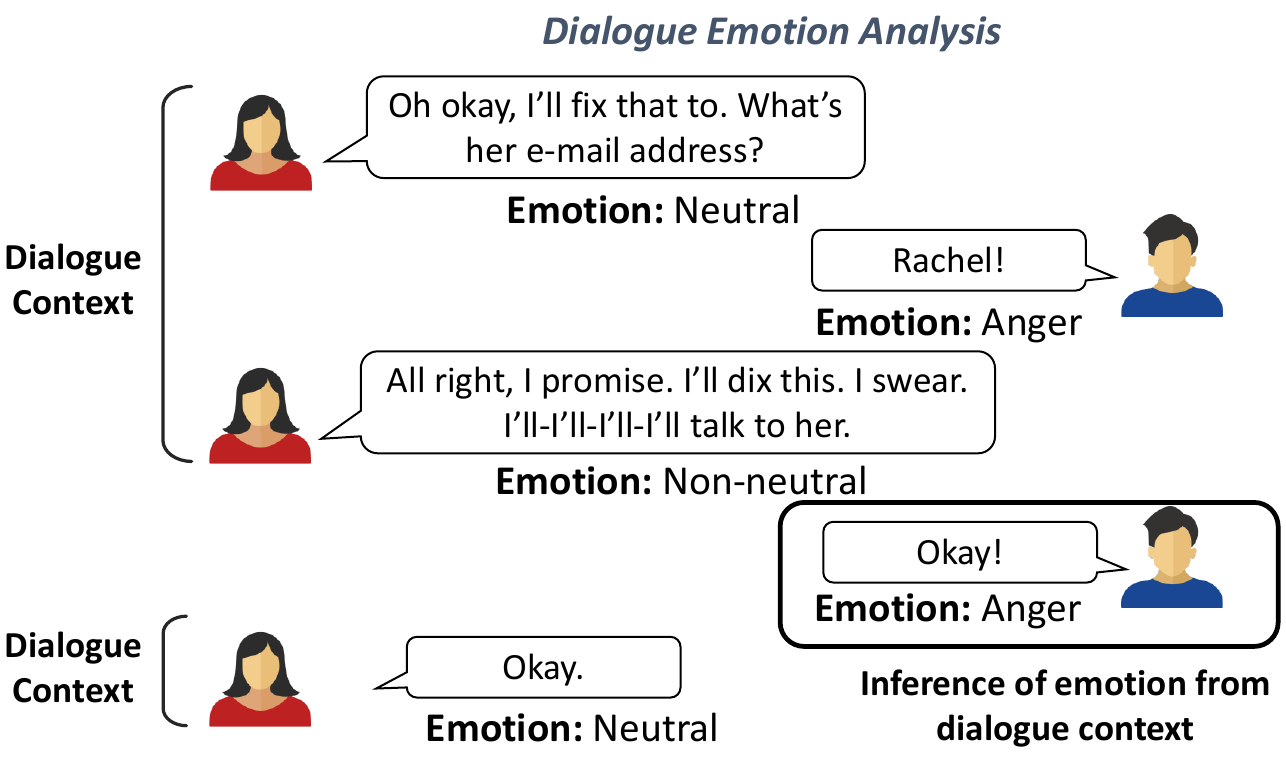}}
	\end{minipage}
	\vspace{-0.6cm}
	\caption{Examples of dialogue emotion analysis from the Emotionlines dataset \cite{hsu2018emotionlines}. Dialogue emotion analysis is about contextual understanding of emotion conveyed in an utterance with the help of the dialogue context. }
	\label{figure_dialogue_emotion}
	\vspace{-0.30cm}
\end{figure}
\subsection{Dialogue Emotion Analysis}

\subsubsection{\textbf{Social Background}} Dialogue emotion analysis entails emotion detection and sentiment analysis in conversations. It refers to the process of recognizing and classifying the types of emotions or sentiments expressed in dialogues. Its aim is to identify the underlying emotional tone (\emph{e.g.}, happy, angry, sad) or sentiment polarity (\emph{e.g.}, positive, neutral, negative). 

Emotion is inherent to human beings and it is closely related with humans' thoughts, feelings and behaviors. The ability to recognize and demonstrate emotions such as fear has started to emerge in the infant stage \cite{lewis1974self}. At approximately two years old, children begin to use words to express their emotions \cite{bloom1998language}. Natural language is used by individuals to reflect their emotions and as a result, dialogue emotion analysis is essential for artificial intelligence to learn social cues and reason about humans' mental states and unobservable beliefs\cite{schwarz2015language, wu2018rational}. Dialogue emotion analysis also receives much attention in academia given its potential use in conversational systems to produce emotion-aware and empathetic dialogues \cite{poria2019emotion}. 

\subsubsection{\textbf{Dialogue Emotion Recognition}} Based on Poria \emph{et al.} \cite{poria2019emotion}, the essential information required to accurately detect the emotion in an utterance from a dialogue can be categorized into three types: 1) the utterance and its surrounding context; 2) the speaker's state; and 3) emotions in preceding utterances. 

Revolving around these three identified categories, researchers have come up with different methods to identify emotions embedded in conversations. Poria \emph{et al.} applied bidirectional LSTM and attention mechanism to model context from surrounding utterances and Hazarika \emph{et al.} introduced conversational memory network to exploit speaker memories to model the dialogue context \cite{hazarika2018conversational}.
These early works did not differentiate between various speakers. In light of this, Majumder \emph{et al.} proposed DialogueRNN \cite{majumder2019dialoguernn}, which treats each party involved in a dialogue individually and adapts to the speaker of every utterance. Li \emph{et al.} \cite{li2020multi} framed the task under a multi-task setting by introducing an auxiliary task of speaker identification to improve emotion recognition capability in models. Besides individual speaker attributes, modeling inter-speaker relations is also very important. Ghosal \emph{et al.} \cite{ghosal2019dialoguegcn} introduced DialogueGCN with a graph network to model inter-speaker dependency. On top of inter-speaker relation modeling, researchers also use graph neutral networks to model dialogue context such as local and global context \cite{sheng2020summarize}, as well as turn modeling \cite{lee2021graph}. Furthermore, some work incorporated common sense knowledge to enhance emotion recognition in dialogues. Ghosal \emph{et al.} \cite{ghosal2020cosmic} introduced the COSMIC framework to incorporate common sense knowledge such as causal relations to better understand interactions and emotion tones. Xie \emph{et al.} \cite{xie2021knowledge} introduced Knowledge-Interactive network which utilizes common sense knowledge and sentiment lexison to enrich semantic information. More recently, Lei \emph{et al.} \cite{lei2023instructerc} framed emotion recognition in conversation to a generative plug-and-play framework using Large Language Models, which achieves state-of-the-art performance.

\subsubsection{\textbf{Discussion}} Past research has made significant progress in dialogue emotion analysis. However, there still exist many research challenges \cite{poria2019emotion, fu2023emotion}. When researchers choose a simpler categorization model with fewer types defined, they are unable to capture fine-grained and nuanced emotions. On the other hand, a complex taxonomy may result in high difficulty in differentiating highly similar emotions and lead to disagreements among annotators. Second, there may exist gaps between intended emotion from the speaker and perceived emotion by the annotator, leading to bias and inaccuracies in annotations. In addition, the emotional state of a speaker can be influenced by various personal factors such as personality and individual beliefs. These speaker-specific variables add another layer of complexity to emotion detection. Emotions are expressed and perceived differently across various social and cultural contexts. Therefore, those social and cultural contexts also should be considered.


\begin{figure}[t]
	\begin{minipage}[b]{1.0\linewidth}
		\centering
		\centerline{\includegraphics[width=9.0cm]{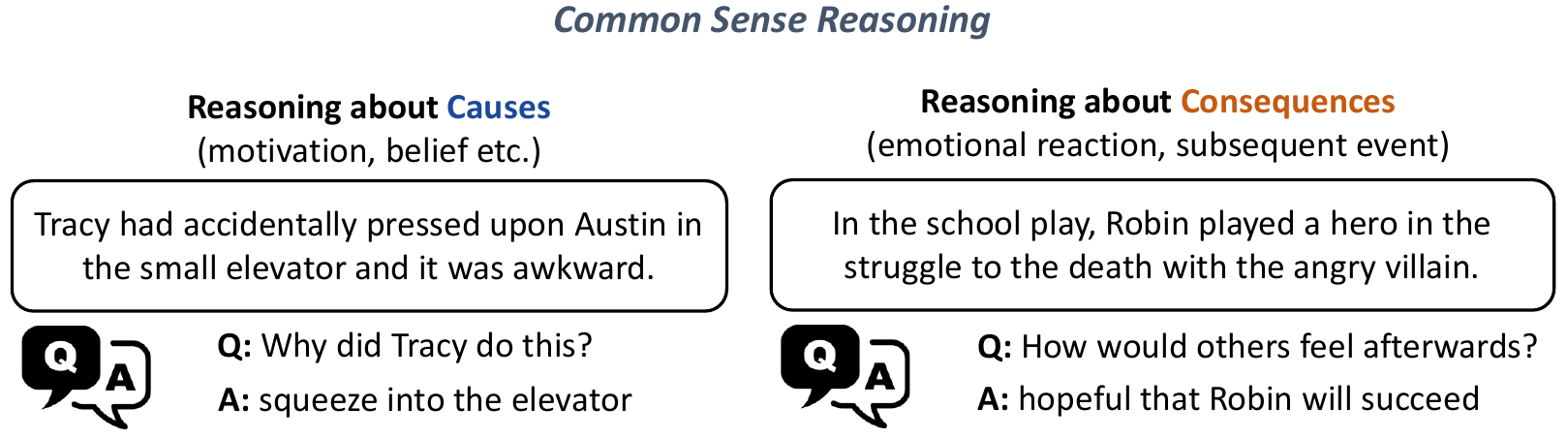}}
	\end{minipage}
	\vspace{-0.6cm}
	\caption{Examples of common sense reasoning from the SocialIQA dataset \cite{sap2019social}. Common sense reasoning includes understanding the causes and consequences of social events.}
	\label{figure_common_sense}
	\vspace{-0.30cm}
\end{figure}
\subsection{Common Sense Reasoning}

\subsubsection{\textbf{Social Background}} Common sense refers to the understanding of the world and common sense reasoning is the capability in human beings to draw logical conclusions, formulate judgments and make decisions based on their understanding. 

Common sense reasoning helps to explain the mechanisms of human mind and behavior \cite{burke1969grammar, moore2013development}. It also aids individuals explore the surroundings and understand the social context \cite{sap2020commonsense}. During social interactions, common sense reasoning allows individuals to make accurate inferences about others' beliefs based on observable evidence and social context. Ambiguity in language remains an important challenge for language understanding and common sense reasoning is one key in resolving such ambiguity \cite{davis2015commonsense}.

\subsubsection{\textbf{Common Sense Reasoning}} Researchers have focused on various tasks to benchmark common sense reasoning capabilities in language models. These tasks include coreference resolution where models need to resolve ambiguities in text; question answering which requires models to understand the context and provide accurate responses; textual entailment where models must infer the relationship between two sentences; plausible inference which tests the model's ability to infer plausible subsequent events using common sense; and psychological reasoning which involves understanding emotions, intentions, and beliefs. Some comprehensive benchmarks cover a wide range of reasoning tasks.

In the past decade, neural methods have demonstrated remarkable progress in common sense reasoning tasks. In early work, embedding models such as word2vec \cite{mikolov2013distributed} and GloVe \cite{pennington2014glove} are trained to learn representations of different words and entities in the vector space. Such word representations can help models understand the relationships between different concepts, enhancing their common sense reasoning capabilities. Model architectures like memory networks \cite{weston2015memory} and Recurrent Entity Network \cite{henaff2016tracking} can keep track of the world state with long-term memory. In more recent years, the transformer architecture has been applied in
the Generative Pre-trained Transformer (GPT) \cite{radford2018improving} and the Bidirectional Encoder Representations from Transformers (BERT) \cite{kenton2019bert}. With such contextual language models, the meaning of a word can be different and hence is better captured based on the surrounding context, leading to stronger common sense reasoning capabilities in models. These pretrained models can be further fine-tuned for downstream common sense reasoning tasks to become more task-specific \cite{xu2022human}. With increased scale in language models, recent large language models can perform common sense reasoning in zero-shot or few-shot \cite{zelikman2022star} manners. Moreover, Wei \emph{et al.} introduced step-by-step thinking through Chain of Thought \cite{wei2022chain}, which strengthens models' common sense reasoning performance. On top of learning from training data used in pretrained language models, researchers also incorporate external knowledge to improve models' common sense reasoning ability. For example, Lv \emph{et al.} \cite{lv2020graph} leveraged heterogeneous knowledge bases like ConceptNet and Wikipedia and built knowledge graphs to learn graph-based contextual word representation. 

Theory of Mind (ToM) \cite{wellman1992child} is a crucial aspect of reasoning in social contexts. ToM refers to the ability to attribute mental states—beliefs, intentions, desires, emotions—to others and to understand that these mental states may differ from one's own. This cognitive capacity is fundamental for social interaction and communication. Baker \emph{et al.} \cite{baker2011bayesian} proposed a Bayesian Theory of Mind (BToM) framework that models joint inference of beliefs and desires, demonstrating how people can rationally infer others' mental states from observed actions. Sap \emph{et al.} \cite{sap2019atomic} developed ATOMIC, a commonsense knowledge graph that captures social and psychological aspects of everyday events, including people's intents and reactions. Another work \cite{rashkin2018event2mind} proposed a neural network approach to model people's mental states from textual narratives. They demonstrated that their model could improve the understanding of social phenomena in stories. These works showcase the potential of integrating Theory of Mind into AI for better social understanding.

\subsubsection{\textbf{Discussion}} Common sense reasoning can be the basis for understanding and navigating social interactions. Despite the advancements made in the field of common sense reasoning, there still remain some challenges. The first challenge lies in the difficulty to achieve human-level common sense reasoning by only using limited training data to train models \cite{storks2019recent}. Although external knowledge bases can be used, data sparsity issue still exists as the knowledge bases are non-exhaustive and may not contain all the knowledge required for a specific reasoning task \cite{cambria2011isanette}. Furthermore, when learning the relationships between entities and concepts, models may be susceptible to spuriousness issue and learn the shortcuts instead of robust generalizations \cite{branco2021shortcutted}.

\begin{table*}[t]
\center
\caption{Datasets for understanding verbal social cues. The table provides attributes of datasets for dialogue act analysis, dialogue emotion analysis, and common sense reasoning tasks.
}
\vspace{-0.2cm}
\small
\renewcommand{\arraystretch}{0.8}
\renewcommand{\tabcolsep}{1.5mm}
\resizebox{\linewidth}{!}{
\begin{tabular}{c  c  c  c  c  c }
\toprule 
\makecell[l]{Dataset} & {Year} & {Data Types} & {Label Types} & {Data Size}   & {Remarks} \\
\midrule
\multicolumn{6}{c}{\bf \makecell{Dialogue Act Analysis}} \\ \midrule

\makecell[l]{SwitchboardDA~\cite{stolcke2000dialogue}} & 2000 & Audio, Text & Around 60 speech tags & 1,155 Conversations  & Telephone call \\
\makecell[l]{MRDA Corpus~\cite{shriberg2004icsi}} & 2004 & Audio & 39 Specific dialogue act tags & 72 Hour meetings  & Group meeting, Multi-party \\
\makecell[l]{AMI Corpus~\cite{kraaij2005ami}} & 2005 & Video & 15 Dialogue act tags & 100 Hour meetings  &  Group meeting, Multi-party  \\
\makecell[l]{DSTC~\cite{Williams2013TheDS}} & 2013 & Audio, Text & Varying number of dialogue states & 1,500 Conversations  &  Telephone call\\
\makecell[l]{DSTC2~\cite{henderson2014second}} & 2014 & Audio, Text & Varying number of dialogue states & 3,235 Conversations  &  Telephone call  \\
\makecell[l]{DSTC3~\cite{henderson2014third}} & 2014 & Audio, Text & Varying number of dialogue states & 2,275 Conversations & Telephone call  \\
\makecell[l]{HOPE~\cite{malhotra2022speaker}} & 2022 & Text & 12 Dialogue acts & 12,900 Utterances  &  Counselling session  \\
\makecell[l]{IMCS-21~\cite{chen2023benchmark}} & 2023 & Text & 8 Specific dialogue act tags & 4,116 Dialogues  &  Medical consultation \\
\midrule
\multicolumn{6}{c}{\bf \makecell{Dialogue Emotion Analysis}} \\ \midrule

\makecell[l]{DailyDialogue \cite{li2017dailydialog}} & 2017 & Text & 6 Emotion classes & 13,118 Dialogues  & Daily dialogue, Multi-turn \\
\makecell[l]{EmotionLines \cite{hsu2018emotionlines}} & 2018 & Text & 7 Emotion classes & 2,000 Dialogues  & TV-show \& Messenger, Multi-party \\
\makecell[l]{EmoContext \cite{chatterjee2019semeval}} & 2019 & Text & 4 Emotion classes & 38,424 Dialogues  & User-Agent interaction \\
\makecell[l]{EmoWOZ \cite{feng2022emowoz}} & 2022 & Text & 7 Emotion classes & 11,434 Dialogues  & User-Agent interaction, Multi-turn \\

\midrule
\multicolumn{6}{c}{\bf \makecell{Common Sense Reasoning}} \\ \midrule
\makecell[l]{WS Challenge~\cite{levesque2012winograd}} & 2012 & Text & 2 Disambiguation options to choose & 273 Sentences  &  Coreference resolution \\
\makecell[l]{SNLI~\cite{bowman2015snli}} & 2015 & Text & 3 Types of logical connections & 570,000 Pairs  &  Text entailment \\
\makecell[l]{Story Cloze Test~\cite{mostafazadeh2016corpus}} & 2016 & Text & 2 Endings and label indicating the correct ending & 3742 Stories  &  Stories ending inference\\
\makecell[l]{JOCI~\cite{zhang2017ordinal}} & 2017 & Text & 5 Plausibility levels & 39,093 Pairs  &  Psychological inference \\
\makecell[l]{Story Commonsense~\cite{rashkin2018modeling}} & 2018 & Text & 5 Motivation classes \& 8 emotion classes & 15,000 Stories  &  Psychological inference \\
\makecell[l]{Event2Mind~\cite{rashkin2018event2mind}} & 2018 & Text & People's possible intents \& reactions & 24,716 Events  &  Psychological inference \\
\makecell[l]{CommonsenseQA~\cite{talmor2019commonsenseqa}} & 2019 & Text & MCQ questions & 12,247 Questions  & Question Answering \\
\makecell[l]{SocialIQA~\cite{sap2019social}} & 2019 & Text & MCQ questions & 38,000 Questions  &  Question Answering \\

\bottomrule 
\end{tabular}
}
\vspace{-0.4cm}
\label{table_dataset_verbal}
\end{table*}

\subsection{Datasets with Verbal Cues}
Table \ref{table_dataset_verbal} shows a comprehensive overview, comparing the attributes of datasets with verbal social cues. Further details for each dataset are included in the supplementary material.

\section{Understanding Non-verbal Cues}

\begin{figure}[t]
	\begin{minipage}[b]{1.0\linewidth}
		\centering
		\centerline{\includegraphics[width=9.0cm]{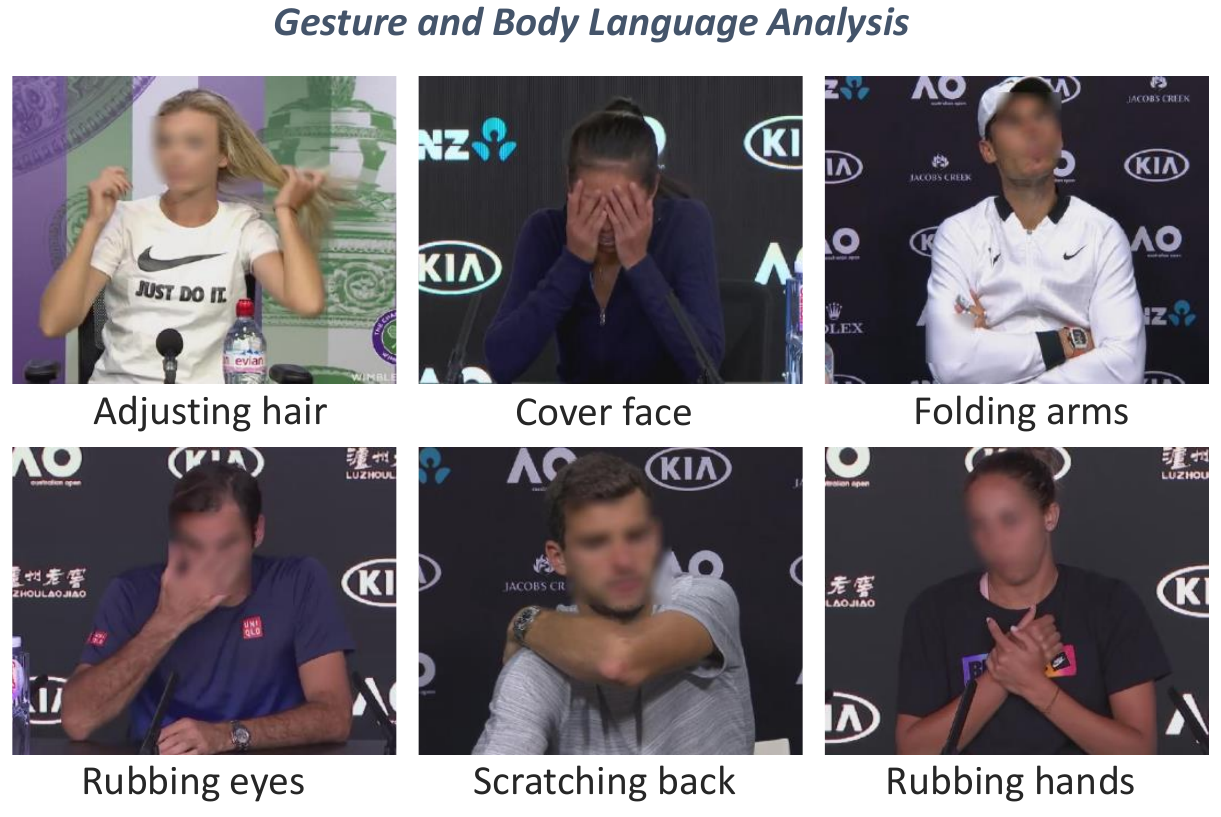}}
	\end{minipage}
	\vspace{-0.6cm}
	\caption{Examples of gesture and body language from iMiGUE~\cite{liu2021imigue}. People convey subtle social nuances through hand, head, and body movements.}
	\label{figure_gesture}
	\vspace{-0.30cm}
\end{figure}

\subsection{Gesture and Body Language Analysis}

\subsubsection{\textbf{Social Background}} When people engage in face-to-face communication, hand movements often accompany their words to emphasize points or release tension. These instinctive hand motions, synchronized with speech, are termed co-speech gestures, and they naturally enhance languages. Across different cultures and linguistic backgrounds, gestures and body language are integral to communication~\cite{clough2020role,goldin1999role}. Notably, babies gesture even before uttering their first words~\cite{kuhn2014early}. Theoretical models posit that both speech and gesture spring from a shared representational system~\cite{mcneill2012language,kobsa1986combining}. This perspective suggests that comprehensive conceptual frameworks encompass both imagery-driven and symbolic data, such as gestures and speech, respectively~\cite{quek2002multimodal,mcneill2000language}. McNeill \textit{et al.}~\cite{mcneill1992hand} identified four types of gesture: deictic (\textit{e.g.}, pointing), beat (\textit{e.g.}, rhythmic emphasis), iconic (\textit{e.g.}, mimicking form), and metaphoric (\textit{e.g.}, abstract).

\subsubsection{\textbf{Isolated Gesture Recognition}} Early works in gesture recognition mainly addressed isolated gestures, which are standalone hand motions or signs. They are identified as separate entities, without considering a sequence of multiple gestures. Early works in isolated gesture recognition utilized optical flow and human pose analysis. The emergence of CNN and 3D CNN has made certain breakthroughs in this field~\cite{wang2016large,zhu2016large}. Lee \emph{et al.}~\cite{lee2018motion} utilized CNN extract feature maps from static images and then proposed a motion filter to merge 2-frame motion information to predict isolated gestures. Zhou \emph{et al.}~\cite{zhou2018temporal} further proposed the framework of temporal relation networks for utilizing temporal relations among multiple CNN-based video frames, extending the idea of relation network~\cite{santoro2017simple} into gesture recognition. Recent work~\cite{li2023learning} attempted to mitigate the impact of gesture-irrelevant information (\emph{e.g.}, background, illumination) by introducing information-theoretic loss for isolated gesture recognition. Research on 3D skeleton gesture recognition focusing on isolated gestures has also gained considerable attention, as demonstrated by the works of De Smedt \emph{et al.}~\cite{de2016skeleton}, who utilized Fisher vector representation with Gaussian Mixture Models. Additionally, Aich \emph{et al.}~\cite{aich2023data} have proposed a strategy by developing a boundary-aware prototypical sampling technique. This method improves model inversion for class-incremental learning in 3D skeleton gesture recognition, notably without relying on pre-existing data.

\subsubsection{\textbf{Continuous Gesture Recognition}} However, in real-world environments, each video clip doesn't contain just one gesture and there can be multiple gestures in the continuous video stream. Therefore, we also need to figure out when each gesture starts and ends simultaneously. Building on the foundation established by research in isolated gesture recognition, early endeavors in continuous gesture recognition aimed to build multi-stage pipelines to separate gesture sequences into distinct isolated gesture features for classification. Notable methods include employing a two-stream RNN~\cite{chai2016two} or 3D CNN~\cite{camgoz2016using} for classification, and leveraging 3D CNNs to derive spatial feature maps subsequently integrated with LSTM for analysis~\cite{cao2017egocentric}. Bhatnagar \emph{et al.}~\cite{bhatnagar2023long} further applied the dynamic neural network to select features for 3D CNN to recognize continuous gestures. In addition, extracting additional features from depth maps within RGB-D videos, as well as analyzing human poses or facial expressions, can also significantly enhance the performance of continuous gesture recognition. In this area, Zhou \emph{et al.}~\cite{zhou2020spatial} introduced a spatial-temporal multi-cue network, a sophisticated approach that amalgamates various features including human poses, gestures, and facial expressions. Besides, Liu \emph{et al.}~\cite{liu2022ld} developed a two-stream CNN framework that utilizes RGB-D video inputs to accurately detect and classify continuous gestures, showing the potential of combining depth maps for continuous gesture recognition.

\subsubsection{\textbf{Discussion}} Gestures and body language play a crucial role in conveying social cues and regulating the flow of interactions. Accurate gesture recognition is essential for understanding the nonverbal aspects of social communication. However, current gesture recognition benchmarks have several limitations in capturing the complexity of social interactions. Most existing benchmarks lack communicative social gestures, multi-person scenarios, and cultural variations in nonverbal behaviors. As a result, most gesture recognition models are limited to simple signal communication tasks like sign language recognition and human-machine interaction. Future work should focus on creating more comprehensive and culturally diverse datasets that can handle the contextual and interpersonal dynamics of gestures in social interactions.

\subsection{Gaze and Attention Analysis}

\begin{figure}[t]
	\begin{minipage}[b]{1.0\linewidth}
		\centering
		\centerline{\includegraphics[width=9.0cm]{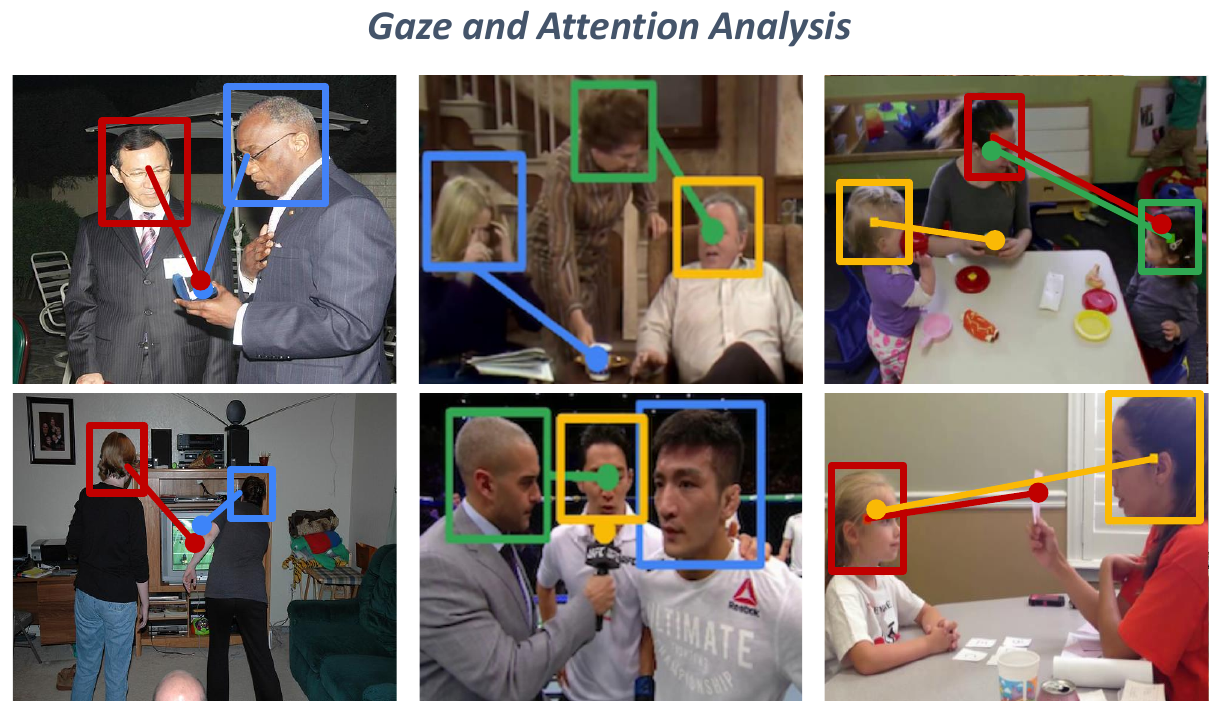}}
	\end{minipage}
	\vspace{-0.6cm}
	\caption{Examples of human gaze from GazeFollow~\cite{recasens2015they}, VideoAttentionTarget~\cite{chong2020detecting}, and ChildPlay~\cite{tafasca2023childplay}. By analyzing human gaze, it is possible to find out their areas of interest and clarify their social intention.}
	\label{figure_gaze}
	\vspace{-0.30cm}
\end{figure}

\subsubsection{\textbf{Social Background}} Gaze behavior plays an important role in understanding social interactions. Where a person looks provides us a lens into how they perceive a scene - for instance, who they are listening to in a conversation, or what stimuli they are paying attention to. Likewise, gaze reflects a person's intent: a speaker often looks towards the target of their speech, or looks at the object to which they are referring. Even before gaining language skills, children learn to use gaze to facilitate communicative processes like joint attention joint attention to socially engage with caretakers \cite{tomasello1986joint}. As language is acquired, eye gaze plays an important role in regulating multi-person conversational processes like turn-taking \cite{kendon1967some}. In this section, we review computer vision progress towards identifying targets of gaze, and using gaze behavior to recognize higher-level social behaviors.

\subsubsection{\textbf{Gaze Target Detection}} Several works have addressed the task of gaze target detection, which aims to localize the target of a person's visual attention in an image containing the person and the scene. Early works in this direction combined head-tracking with estimation of head orientation to identify when a person is looking at other people in meetings \cite{stiefelhagen1999modeling, stiefelhagen2001estimating}. Building upon a larger body of work in visual saliency prediction \cite{ullah2020brief}, Recasens \emph{et al.} \cite{recasens2015they} published a seminal work introducing the GazeFollow benchmark. They proposed a 2-stream deep neural network with separate branches to estimate gaze direction and scene saliency, combining their outputs to produce a gaze heatmap over the scene for a given person. Several works have improved upon this 2-stream paradigm \cite{saran2018human, chong2017detecting, lian2018believe, zhao2020learning, chen2021gaze, wang2022gatector}, including adapting to 360-degree images \cite{li2021looking}. Notably, Chong \emph{et al.} \cite{chong2020detecting} introduced the VideoAttentionTarget dataset, which provides gaze target annotations for video clips. They proposed a 2-stream model with an Convolutional LSTM \cite{shi2015convolutional} to integrate temporal information. Many recent works have demonstrated performance gains by augmenting 2-stream models with pretrained models for estimating additional signals like scene depth \cite{fang2021dual, bao2022escnet, tafasca2023childplay}, body pose \cite{bao2022escnet, gupta2022modular}, head pose \cite{fang2021dual, horanyi2023they}, and eye detection \cite{fang2021dual}. A set of recent works has followed a detection framing of the problem, where a set of head bounding boxes and paired gaze targets are detected end-to-end via a DETR-based \cite{carion2020end} Transformer network \cite{tu2022end, tonini2022multimodal, tonini2023object}. 

Other works predict gaze targets from the egocentric point of view. These methods use video from a headworn camera to predict the gaze of the camera wearer. This problem also differs from eye-tracking, which uses close-up cameras capturing the eyes to determine gaze direction; rather, gaze is localized from the egocentric video stream, which captures the scene. Fathi \emph{et al.} \cite{fathi2012learning} introduced the GTEA and GTEA+ datasets of daily activities with ground truth eye-tracking data, and proposed a probabilistic model for inferring gaze conditioned on action. Further methods predict gaze based on saliency, hand, and egomotion features \cite{yamada2011attention, li2013learning}. Huang \emph{et al.} and Li \emph{et al.} proposed deep learning methods for jointly modeling visual attention and activity \cite{huang2018predicting, huang2020mutual, li2021eye}. Lai \emph{et al.} \cite{lai2022eye,lai2023eye} proposed a Transformer-based model and further evaluated on the Ego4D \cite{grauman2022ego4d} eye-tracking subset, which focuses on multi-party social interactions. A related task is \textit{anticipating} egocentric gaze in future frames \cite{zhang2017deep, zhang2018anticipating, lai2023listen}.

\subsubsection{\textbf{Multi-party Gaze Analysis}} A lesser studied direction of gaze analysis has focused on recognizing multi-person social gaze dynamics. In contrast to gaze target detection, tasks in this direction involve reasoning at a higher-level about people and their respective gaze targets in relation to others. One such problem is recognizing shared attention, or when 2 people in a scene are looking at the same target \cite{hoffman2006probabilistic, fan2018inferring, sumer2020attention, nakatani2023interaction}. Park \emph{et al.} defined the related task of social saliency prediction, which involves identifying potential targets of shared attention in scenes with multiple people \cite{park20123d, park2013predicting, soo2015social}. They propose using multiple first-person videos to reconstruct the scene and learn a social saliency model that is applied downstream to third-person images. Prior works have also addressed the task of determining when people look at eachother, referred to as mutual gaze detection or ``Looking at Eachother" (LAEO) \cite{marin2011here, marin2014detecting, palmero2018automatic, marin2019laeo, doosti2021boosting, medina2021suarez, guo2022mgtr}. Muller \emph{et al.} propose using gaze and speaking behavior together to identify eye-contact in multi-party interactions \cite{muller2018robust}. In 2019, Fan \emph{et al.} \cite{fan2019understanding} introduced the VACATION benchmark  with various types of gaze behaviors such as shared attention, mutual gaze and gaze following. They proposed a spatio-temporal graph approach for capturing gaze dynamics between people over time. Further works have developed end-to-end Transformer-based architectures for identifying these gaze behaviors \cite{de2023temporal, nakatani2023interaction}. 

Researchers have also addressed identifying social gaze dynamics in egocentric videos. Works have studied the task of detecting when a person visible in the egocentric video stream is looking at the camera wearer. Methods have addressed eye-contact detection \cite{ye2012detecting, smith2013gaze, ye2015detecting, mitsuzumi2017deep, zhang2017everyday, watanabe2019spatio, chong2020detection}, with several focusing on the use-case of assessing child behavior in the clinical setting \cite{ye2012detecting, ye2015detecting, chong2020detection}. Similarly, the Ego4D dataset \cite{grauman2022ego4d} includes the ``Looking at Me" task, and the related audiovisual ``Talking to Me" task as part of their social benchmark. Paralleling the task of identifying shared attention in third-person videos, researchers have explored identifying objects of shared attention across multiple egocentric videos by identifying similarities in views \cite{kera2016discovering, huang2017temporal, huang2020ego}. Fathi \emph{et al.} \cite{fathi2012social} addressed identifying social interaction types via inferring the visual attention of both the camera-wearer and people in their view, including dialogue, monologue, and discussion .

\subsubsection{\textbf{Discussion}} Gaze and attention are important indicators in human interactions. The works on recognizing multi-party gaze dynamics represent an important step towards understanding the dimension of gaze in social interactions. However, there is relatively limited work that addresses gaze in the context of other modalities and social cues \cite{muller2018robust, grauman2022ego4d, lai2023listen}. The joint modeling of gaze with verbal and non-verbal cues is crucial for a holistic understanding of social communication. We see the joint modeling of gaze along with verbal and other nonverbal cues as an important and necessary direction towards holistic understanding of social interactions. Moreover, current research on gaze often does not consider the contextual factors that influence gaze behaviors.

\subsection{Facial Expression Analysis}

\begin{figure}[t]
	\begin{minipage}[b]{1.0\linewidth}
		\centering
		\centerline{\includegraphics[width=9.0cm]{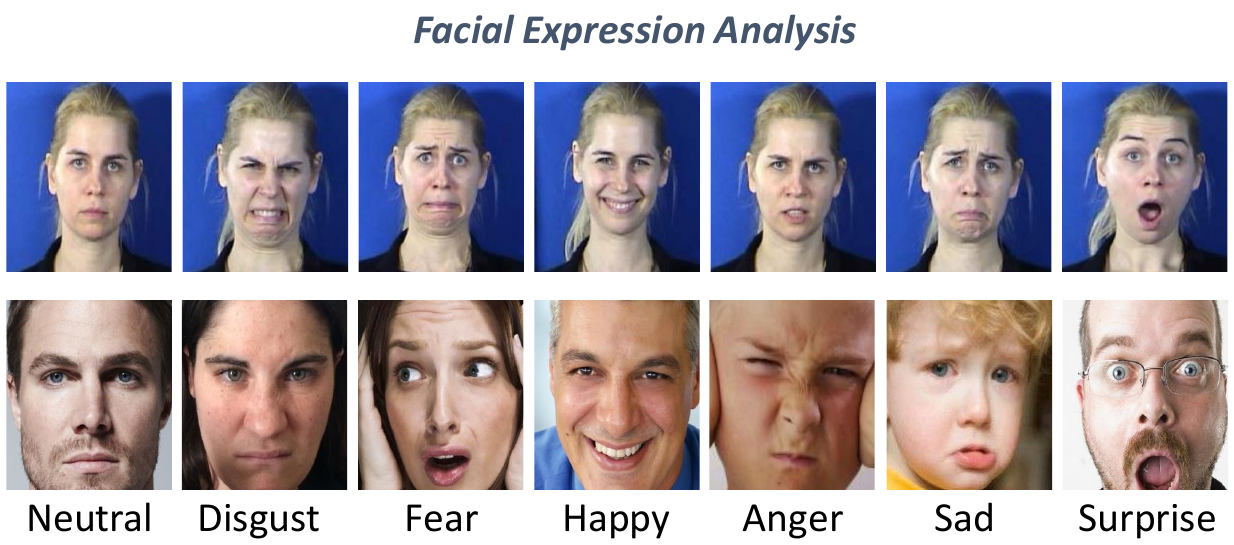}}
	\end{minipage}
	\vspace{-0.6cm}
	\caption{Examples of facial expression from MMI~\cite{pantic2005web,valstar2010induced} and AffectNet~\cite{mollahosseini2017affectnet}. They represent neutral and six core emotions - disgust, fear, happy, anger, sad, and surprise.}
	\label{figure_expression}
	\vspace{-0.30cm}
\end{figure}

\subsubsection{\textbf{Social Background}} Facial expressions serve as a vital channel for humans to communicate their emotions, playing a key role in the fabric of social interactions. Facial expressions are more than just transient displays of emotion; they are the silent language through which individuals communicate a wide range of feelings, from joy and surprise to sadness and anger, often without uttering a single word. This section explores advancements in facial expression analysis and underscores its importance in cultivating relationships that are sensitive to emotional and social nuances.

In the field of computer vision, numerous facial expression recognition approaches have been developed to interpret expressions from human faces. The origins of it can be traced back to the 20th century, where Ekman and Friesen~\cite{Ekman1971ConstantsAC} identified six core emotions—anger, disgust, fear, happiness, sadness, and surprise—proposing that these basic emotions are universally recognized across different cultures. Subsequently, contempt was acknowledged as a seventh fundamental emotion, further broadening our comprehension of human emotional expressions~\cite{matsumoto1992more}. Although there are various models for categorizing emotions, such as the facial action coding system ~\cite{ekman1978facial} and the continuous model based on affect dimensions~\cite{gunes2013categorical}, the categorical model is the most widely adopted approach. 

\subsubsection{\textbf{Static Facial Expression Recognition}} In earlier times, traditional approaches~\cite{zhao2007dynamic,zhong2012learning,moore2011local,hu2008multi} relied on manually crafted features~\cite{dalal2005histograms,jabid2010local,ojala2002multiresolution}. However, with the accumulation of sufficient training data since 2013, methods based on deep learning have come into the spotlight, marking a significant shift in the field. In their pioneering work, Savchenko et. al~\cite{savchenko2021facial} demonstrated the potential of CNNs in facial expression recognition. Following this path, Zhao et. al~\cite{zhao2021robust} introduced EfficientFace, where a local-feature extractor and a channel-spatial modulator have been developed to capture local facial characteristics and global-salient features, enhancing the model's resilience to occlusions and variations in pose. However, a notable limitation of convolution-based approaches is their inability to capture the global context of images, owing to the confined scope of the convolutional local receptive field. Addressing this gap, Xue et. al~\cite{xue2021transfer} took inspiration from the vision transformer~\cite{dosovitskiy2020image}, crafting the first transformer-based network. This network was specifically engineered to grasp long-range dependencies within images, marking a significant stride in this field. Building on this advancement, Kim et. al~\cite{kim2022facial} refined the vision transformer (ViT), enhancing its capability to amalgamate both global and local image features. Following prior research, POSTER~\cite{zheng2023poster} tackles the main challenges of facial expression recognition, namely inter-class similarity, intra-class discrepancy, and sensitivity to scale by cleverly integrating facial landmarks with image features using a dual-stream pyramidal cross-fusion transformer architecture. Nonetheless, the computational expense incurred by the network's architecture impacts the efficiency of facial expression recognition. To mitigate this problem, POSTER++~\cite{mao2023poster} has been introduced. This approach reconstructs the network structure by streamlining the cross-fusion and multi-scale extraction processes, achieving state-of-the-art performance.

\subsubsection{\textbf{Dynamic Facial Expression Recognition}} Advancements in static facial expression recognition (SFER) are now being complemented by growing interest in dynamic facial expression recognition (DFER), which adds the complexity of analyzing temporal alongside spatial features in images.
Several techniques\cite{lu2018multiple,zhang2018facial} leverage CNNs to extract spatial features from frames, then apply RNNs for temporal analysis. 3D CNNs are also used as an alternative. Recently, transformers have emerged as effective tools for extracting both spatial and temporal data, with Zha \emph{et al.}'s~\cite{zhao2021former} dynamic facial expression recognition transformer (Former-DFER) incorporating a convolutional spatial transformer and a temporal transformer. Ma \emph{et al.}~\cite{ma2022spatio} introduced the spatial-temporal Transformer (STT) for capturing frame-specific features and their contextual relationships. Additionally, Wang \emph{et al.}~\cite{wang2022dpcnet} developed the Dual Path multi-excitation Collaborative Network (DPCNet) to learn facial expression features from key frames.

\subsubsection{\textbf{Discussion}} Facial expressions are a key channel for conveying emotions and social signals in human interactions. However, current facial recognition systems often struggle with biases related to gender, race, and cultural background \cite{Steephen2018DoWE, buolamwini2018gender, Klare2012Face, Chen2021Understanding}. These biases can lead to the misinterpretation or misuse of facial expression analysis in social contexts, perpetuating stereotypes and discrimination against underrepresented groups. While various approaches have been proposed to mitigate these biases, effectively addressing them remains a significant challenge \cite{Robinson_2020_CVPR_Workshops,wang2020towards,xu2020investigating,Churamani2022DICL4Bias,Kara2021Towards}. Future research should develop more inclusive and unbiased facial expression recognition systems. In addition, privacy is an important issue in systems that handle facial data. The use of these systems without proper regulation can lead to the violation of privacy rights and the misuse of their personal information \cite{zhang2021facial}. Therefore, it is also important to develop AI systems that consider privacy protection and ensure the responsible use of facial data.

\begin{table*}[t]
\center
\caption{Datasets for understanding non-verbal social cues. The table provides attributes of datasets for gesture and body language analysis, gaze and attention analysis, and facial expression tasks.
}
\vspace{-0.2cm}
\small
\renewcommand{\arraystretch}{0.8}
\renewcommand{\tabcolsep}{0.2mm}
\resizebox{\linewidth}{!}{
\begin{tabular}{ c  c  c  c  c  c  }
\toprule 
\makecell[l]{Dataset} & {Year} & {Data Types} & {Label Types} & {Data Size}  & {Remarks} \\
\midrule 
\multicolumn{6}{c}{\bf \makecell{Gesture and Body Language Analysis}} \\ \midrule
\makecell[l]{ChaLearn IsoGD~\cite{wan2016chalearn}} & 2016 & Video & 249 Gesture classes & 47,933 Videos & Lab, Isolated  \\
\makecell[l]{ChaLearn ConGD~\cite{wan2016chalearn}} & 2016 & Video & 249 Gesture classes & 22,535 Videos & Lab, Continuous \\
\makecell[l]{EgoGesture~\cite{zhang2018egogesture}} & 2018 & Video & 83 Gesture classes & 2,081 Videos (27h) &  Lab, Egocentric, Isolated  \\
\makecell[l]{Jester~\cite{materzynska2019jester}} & 2019 & Video & 27 Gesture classes & 148,092 Videos (123h) & Lab, Webcam, Isolated \\
\makecell[l]{IPN Hand~\cite{benitez2021ipn}} & 2020 & Video & 13 Gesture classes & 200 Videos (7h) &  Lab, Isolated  \\
\makecell[l]{iMiGUE~\cite{liu2021imigue}} & 2021 & Video & 32 Gesture classes & 359 Videos (35h) &  Web, Interview, Isolated  \\
\makecell[l]{LD-ConGR~\cite{liu2022ld}} & 2022 & Video & 10 Gesture classes & 542 Videos &  Lab, Continuous  \\
\makecell[l]{SMG~\cite{chen2023smg}} & 2023 & Video & 17 Gesture classes & 40 Videos (8h) &  Lab, Isolated \& Continuous  \\
\makecell[l]{HaGRID~\cite{kapitanov2024hagrid}} & 2024 & Image & 18 Gesture classes & 552,992 Images &  Lab, Isolated  \\
\midrule 
\multicolumn{6}{c}{\bf \makecell{Gaze and Attention Analysis}} \\ \midrule
\makecell[l]{GazeFollow~\cite{recasens2015they}} & 2015 & Image & Gaze point & 122,143 Images  & Web, Multi-party \\
\makecell[l]{VideoGaze~\cite{recasens2017following}} & 2017  & Video & Gaze point \& in/out of frame & 140 Videos (166,721 frames) & Movie, Multi-party \\
\makecell[l]{DLGaze~\cite{lian2018believe}} & 2018 & Video & Gaze point & 86 Videos (53min) & Lab, Multi-party \\
\makecell[l]{VideoCoAtt~\cite{fan2018inferring}} & 2018 & Video & Shared attention bounding boxes & 380 Videos (5.5h) & YouTube, Multi-party \\
\makecell[l]{EGTEA Gaze+~\cite{li2018eye}} & 2018 & Video & Eye tracking & 86 Videos (28h) & Lab \\
\makecell[l]{VACATION~\cite{fan2019understanding}} & 2019 & Video & Gazed object boxes \& Gaze classes (6 atomic/5 event) & 300 Videos (1h) & YouTube, Multi-party \\
\makecell[l]{UCO-LAEO~\cite{marin2019laeo}} & 2019 & Video & Mutual for each pair of people & 129 Videos (10min) & TV-show, Multi-party \\
\makecell[l]{AVA-LAEO~\cite{marin2019laeo}} & 2019 & Video & Mutual for each pair of people & 298 Videos (13h) & Movie, Multi-party \\
\makecell[l]{VideoAttentionTarget~\cite{chong2020detecting}} & 2020 & Video & Gaze point + in/out of frame & 606 Videos (71,666 frames) & YouTube, Multi-party \\
\makecell[l]{Triadic Belief~\cite{grauman2022ego4d}} & 2021 & Video & Eye tracking, pointing, events (3 types), beliefs (4 types) & 88 Videos (72m) & Lab, Multi-party \\
\makecell[l]{Ego4D~\cite{grauman2022ego4d}} & 2022 & Video & Eye tracking & 27 Videos (31h) & Lab, Multi-party \\
\makecell[l]{ChildPlay~\cite{tafasca2023childplay}} & 2023 & Video & Gaze point + in/out of frame & 401 Videos (1h) & YouTube, Multi-party \\

\midrule 
\multicolumn{6}{c}{\bf \makecell{Facial Expression Analysis}} \\ \midrule
\makecell[l]{MMI~\cite{pantic2005web,valstar2010induced}} & 2005 & Image, Video & 7 Expression classes & 740 Images, 2,900 Videos &  Lab, Static \& Dynamic \\
\makecell[l]{CK+~\cite{lucey2010extended}} & 2010 & Video & 8 Expression classes & 593 Videos & Lab, Dynamic \\
\makecell[l]{FER-2013~\cite{goodfellow2013challenges}} & 2013 & Image & 7 Expression classes & 35,887 Images &  Web, Static \\
\makecell[l]{SFEW 2.0~\cite{dhall2015video}} & 2015 & Image & 7 Expression classes & 1,766 Images &  Movie, Static \\
\makecell[l]{EmotioNet~\cite{fabian2016emotionet}} & 2016 & Image & 23 Expression classes & 1,000,000 Images &  Web, Static \\
\makecell[l]{AFEW 7.0~\cite{dhall2017individual}} & 2017 & Video & 7 Expression classes & 1,809 Videos &  Movie, Dynamic \\
\makecell[l]{RAF-DB~\cite{li2017reliable,shan2018reliable}} & 2017 & Image & 7 Expressions \& 12 Compound expressions & 29,672 Images &  Web, Static \\
\makecell[l]{AffectNet~\cite{mollahosseini2017affectnet}} & 2017 & Image & 7 Expression classes & 450,000 Images &  Web, Static \\
\makecell[l]{ExpW~\cite{zhang2018facial}} & 2018 & Image & 7 Expression classes & 91,793 Images &  Web, Static \\
\makecell[l]{Face MPMI~\cite{baddar2019mode}} & 2019 & Video & 7 Expression classes & 37,856 Videos &  Lab, Dynamic \\

\bottomrule 
\end{tabular}
}
\vspace{-0.4cm}
\label{table_dataset_nonverbal}
\end{table*}

\subsection{Datasets with Non-verbal Cues}
Table \ref{table_dataset_nonverbal} shows a comprehensive overview, comparing the attributes of datasets with non-verbal cues. Further details for each dataset are included in the supplementary material.

\section{Understanding Multimodal Cues}

\begin{figure}[t]
	\begin{minipage}[b]{1.0\linewidth}
		\centering
		\centerline{\includegraphics[width=9.0cm]{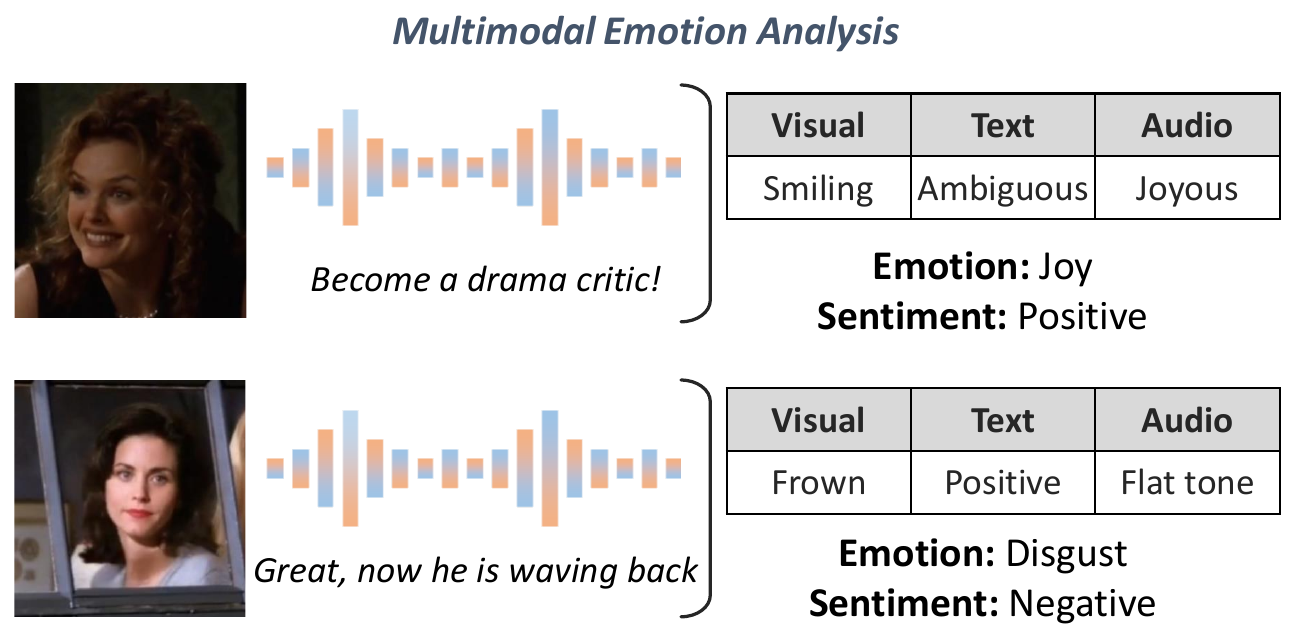}}
	\end{minipage}
	\vspace{-0.6cm}
	\caption{Examples of multimodal emotion and sentiment analysis from MELD \cite{poria2019meld}. By comprehensively considering multimodal verbal and non-verbal social cues, their emotions and sentiments can be interpreted appropriately.}
	\label{figure_multimodal_emotion}
	\vspace{-0.30cm}
\end{figure}

\subsection{Multimodal Emotion Analysis}
\subsubsection{\textbf{Social Background}} Human communication is inherently multimodal, conveying emotions through verbal cues like spoken utterances, as well as non-verbal behaviors including gestures, gaze, and facial expressions. Therefore, it is required to understand multimodal cues for recognizing emotional expressions effectively. For instance, it is difficult to detect sarcasm based purely on utterance texts \cite{castro2019towards}. However, humans can excel at this task by recognizing tone changes and interpreting body language like shrugging shoulders or curling lips. Such multimodal nature adds complexity to emotion analysis \cite{picard2000affective}, and multimodal emotion recognition aims to achieve a more comprehensive and nuanced understanding of human emotional states by leveraging multimodal cues.

\subsubsection{\textbf{Multimodal Emotion Recognition}} Unlike early stages of emotion analysis research, which predominantly focused on single modalities like speech \cite{nicholson2000emotion, el2011survey}, textual content \cite{wu2006emotion, thelwall2010sentiment} and facial expressions \cite{valstar2011first, sariyanidi2014automatic}, the development of multimodal emotion recognition underscores the advantages of jointly considering various channels of expression simultaneously to achieve accurate sentiment recognition\cite{chen2017multimodal, xu2018co, xu2020social}. The emotion recognition in conversation task, contextual information from different modalities becomes particularly indispensable for decoding emotional expression and interpretation \cite{poria2019meld}. Poria \emph{et al.} \cite{poria2017context} propose an LSTM-based model that fuses textual, audio, and visual features to capture interaction history context. Hazarika \emph{et al.} \cite{hazarika2018conversational} introduce a model  that uses two speaker-dependent GRUs as memory to model utterance context. As an extension to CMN, \cite{hazarika2018icon} adds an extra global GRU to track the inter-speaker dependency that exists in the entire conversation. However, neither methods can model multiparty conversations involving more than two speakers because the number of memory network blocks is fixed. Besides, neither of them can distinguish individuals according to given utterances. 

To enable scalability to multi-party datasets and formulate a better context representation, another RNN-based method \cite{majumder2019dialoguernn} employs an attention mechanism. This mechanism considers both the speaker's state and context from preceding states when processing each incoming utterance to infer the current emotion. Recent research has delved into exploring graph convolutional network (GCN) based models to better leverage speaker information to model inter-speaker and intra-speaker dependency. DialogueGCN \cite{ghosal2019dialoguegcn} uses a single textual modality to capture both speaker and conversation sequential information to determine the emotional inertia of individual speakers and the relative positioning of past and future utterances. Subsequent works such as MMGCN \cite{hu2021mmgcn} and MM-DFN \cite{hu2022mm} construct graphs for three modalities—textual, visual, and acoustic data—and leverage these aligned features. M$^{\text{3}}$Net \cite{chen2023multivariate}, explores multivariate information across modalities and context, utilizing multi-frequency signals to examine the importance of emotion discrepancy and commonality. 

\subsubsection{\textbf{Discussion}} Multimodal emotion analysis has made significant progress in recent years, but there are still limitations in both datasets and approaches. The current emotion labels in datasets often hold only a narrow and basic range of emotions, such as happiness or anger, while incorporating fine-grained emotion annotations can enable the recognition of more nuanced emotional states. They can facilitate the study of emotional shifts and capture the dynamic nature of emotional expressions over time. Cultural consideration and individual properties are also under-explored areas. In terms of modeling approaches, current work in multi-party settings primarily relies on holistic visual representations rather than incorporating individual-specific aligned features. Future research can address creating more comprehensive datasets with diverse emotion annotations and developing models that can better capture individual dynamics.

\subsection{Conversation Dynamics Analysis}

\begin{figure}[t]
	\begin{minipage}[b]{1.0\linewidth}
		\centering
		\centerline{\includegraphics[width=9.0cm]{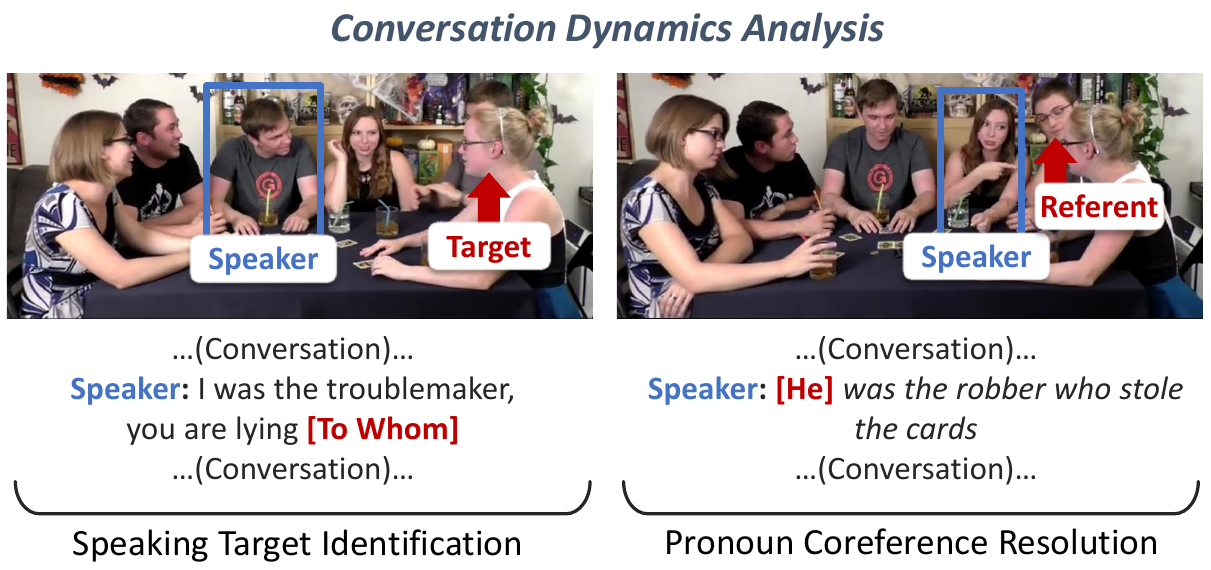}}
	\end{minipage}
	\vspace{-0.6cm}
	\caption{Examples of conversation dynamics analysis in multi-party settings from \cite{lee2024modeling}. Through the consideration of multimodal cues, it is possible to understand the dynamics of the conversation such as who is speaking to whom and who is referred in multi-party environments well.}
	\label{figure_conversation}
	\vspace{-0.30cm}
\end{figure}

\subsubsection{\textbf{Social Background}} Conversation dynamics analysis aims to model the fine-grained interactions among multiple people engaged in a conversation, which is accompanied by modeling the interplay of verbal and non-verbal cues. It involves modeling various aspects of dynamics such as identifying who is speaking to whom, inferring who is the target of attention, and capturing the referential relationships between speakers and listeners. Effective communication requires interpreting these subtle dynamics, which is essential for machines to seamlessly understand multi-party interactions.

\subsubsection{\textbf{Conversation Dynamics Understanding}} To explicitly study gaze behaviors with multimodal cues, Hou \emph{et al.} \cite{hou2024multi} introduced a multimodal gaze following model for conversational scenarios. In the egocentric analysis context, Jiang \emph{et al.} \cite{jiang2022egocentric} involved the multi-channel audio stream to localize the active speaker in all possible directions on a sphere. Ryan \emph{et al.} \cite{ryan2023egocentric} proposed the new task of selective auditory attention localization to detect who the camera wearer is listening to using egocentric videos and multi-channel audios. They introduced a transformer-based framework to model the relationships between audiovisual features in different spatial regions. Jia \emph{et al.} \cite{jia2024audio} introduced the concept of an audio-visual conversational graph to represent the conversational behaviors (speaking and listening) of both the camera wearer and all social partners. Regarding the language-visual modeling, Lee \emph{et al.} \cite{lee2024modeling} tackled the problem of identifying referents in multiparty conversations, curating new social tasks: speaking target identification, pronoun coreference resolution, and mentioned player prediction. They proposed a model that established dense language-visual alignments by synchronizing player visual features with their corresponding utterances. 

Some other works in this area concentrates on modeling the poses and spatial relation of participants in a conversational group, including inferring head, body orientation and facing formations, as well as detecting the independent conversational groups in cocktail party ~\cite{ricci2015uncovering, alameda2015salsa, tan2022conversation, raman2022conflab}. Ricci \emph{et al.} \cite{ricci2015uncovering} proposed a coupled learning framework to jointly estimate head pose, body pose, and F-formations from surveillance videos while handling occlusions. Alameda-Pineda \emph{et al.} \cite{alameda2015salsa} introduced SALSA, which records a social event using cameras and wearable sensors, discussing challenges in crowded interactions. Tan \emph{et al.} \cite{tan2022conversation} presented an LSTM-based model that predicts continuous pairwise affinities to identify conversational groups, leveraging spatial context and temporal dynamics in positional and orientation cues. Joo \emph{et al.} \cite{joo2019towards} introduced a framework to predict social signals such as speaking status, social formation, and body gestures in triadic social interactions based on 3D motion capture data. To this end, they leveraged various types of social cues such as body motions, face expressions, positions, and voice signals.

\subsubsection{\textbf{Discussion}} Current approaches in conversation dynamics analysis primarily focus on integrating raw multimodal signals to identify interaction patterns. 
However, they often lack the incorporation of higher-level knowledge and reasoning about the semantic meaning and intent behind those patterns. Leveraging higher-level knowledge about social norms and cultural practices could enable more intelligent interaction analysis. Future research can explore methods for integrating common sense reasoning, social knowledge, and cultural understanding into conversation dynamics models. This includes developing datasets and benchmarks that capture the diversity of conversation styles across different social contexts and cultural backgrounds, as well as creating models that can adapt to the dynamic and emergent nature of conversation dynamics in real-world social interactions.

\subsection{Social Situation Analysis}

\begin{figure}[t]
	\begin{minipage}[b]{1.0\linewidth}
		\centering
		\centerline{\includegraphics[width=9.0cm]{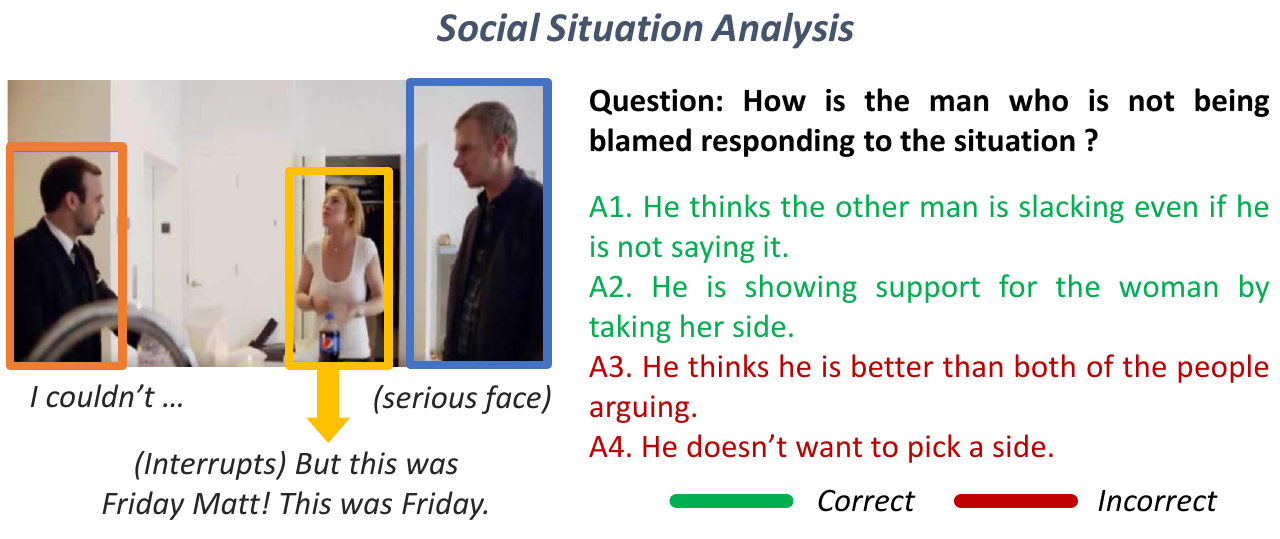}}
	\end{minipage}
	\vspace{-0.6cm}
	\caption{Examples of social situation analysis from Social-IQ \cite{zadeh2019social}. Based on multimodal social cues such as facial expressions and conversation content, social situations can be more accurately interpreted.}
	\label{figure_situation}
	\vspace{-0.30cm}
\end{figure}

\subsubsection{\textbf{Social Background}} In addition to learning each social component separately, human has the ability to understand a social scenario in a holistic way. For example, when two people are arguing on a problem and there's a third person listening to them, we can tell which side the third person is in favor of based on their gestures, facial expressions and utterances. We can also infer infer people's social relationships (\emph{e.g.}, couple, friend, colleague) by observing various cues such as their age, intimacy, and conversations. Such an ability requires fusion of different input modalities (\emph{e.g.}, vision, audio, language, \emph{etc.}), understanding interactions among multiple parties, learning from contexts and involving common sense for social reasoning. We name this capability \emph{social situation analysis}, which is a key step for learning a holistic representation of a specific social scene. We address related works on social relationship reasoning, social question answering, and other tasks related to the analysis of social situations.

\subsubsection{\textbf{Social Relationship Reasoning}} 
Understanding social relationships between people is a key aspect of social situation interpretation. It is also related to understanding beliefs because beliefs about their relationships actually affect the dynamics of social interactions. Early works mainly focused on single modal approaches, analyzing relationships from language cues or visual cues separately. In a language aspect, some works analyzed patterns in dialogue to identify relationships between people, including dialogue-based relation extraction frameworks with the timeliness of communication \cite{ yu2020dialogue }. Other works leveraged graph-based approaches with a multi-view graph to capture various possible relationships \cite{ xue2021gdpnet } and incremental graph parsing \cite{ qiu2021socaog } for inferring social relations. In terms of visual relations, Li \emph{et al.} \cite{ li2017dual} utilized coarse-level and fine-level visual features to infer relationships between people in images, while Sun \emph{et al.} \cite{sun2017domain} leveraged visual attributes such as proximity, pose, and gender to recognize social relations. Other works extended this to generating social relationship graphs from images \cite{goel2019end} and videos \cite{liu2019social}. Recent works have focused on multimodal approaches that combine multimodal cues (\emph{e.g.}, language, visual, audio) for more comprehensive social relationship analysis. Kukleva \emph{et al.} \cite{kukleva2020learning} jointly predicted interactions, social relations, and character pairs using visual and dialog cues, showing that joint learning is effective. Wu \emph{et al.} \cite{wu2021linking} formulated social relation recognition as a social graph generation problem. They integrated short-term multimodal cues to generate frame-level subgraphs and aggregated them to obtain a global video-level social graph for relation reasoning. Wang \emph{et al.} \cite{wang2023shifted} proposed a multimodal graph-based approach for social relation recognition in long videos utilizing shifted graph operations and a memory mechanism.

\subsubsection{\textbf{Social Question Answering}} Social-IQ \cite{zadeh2019social} is a Q\&A benchmark for understanding social situations. It features various in-the-wild social videos, each with questions and multiple-choice answers. The questions address: (1)  causes and intentions behind a social situation, (2) how and why humans act in a certain manner, (3) mental state, trait, attitude, and (4) memory for referencing. The task is to choose the correct answer based on social understanding. Sartzetaki \emph{et at.} \cite{sartzetaki2022extending} developed a model which embeds the visual, textual context and audio inputs with three separate encoders. The question is encoded by a control unit and then injected into the embedding of each modality to learn a question-related representation. Wilf \emph{et al.} \cite{wilf2023face} introduced the graphic model into social intelligence answering. They encode the multimodal input of each speaking turn into a few nodes in a graph and use a factorization node to merge the information. Xie \emph{et al.} \cite{xie2023multi} proposed to leverage emotional characteristics as common sense to select positive and hard negative samples for contrastive learning. They finetune a text encoder on emotion recognition dataset to measure the distance of each answer and the question, which determines the positive and negative pairs for training. Natu \emph{et al.} \cite{natu2023external} used language models to generate some facts for each video based on the transcript. These facts are integrated into the multimodal model as global common sense by straightforward concatenation or treating it as an additional modality. Zhu \emph{et al.} \cite{zhu2023personality} further annotated personality labels in TVQA \cite{lei2018tvqa}. Then they proposed two problems -- predicting (1) the MBTI personality based on the video and (2) the response of a specific character given their personality. 

\subsubsection{\textbf{Other Social Situation Analyses}} In addition to the relationship reasoning and question answering, some work models the social situation analysis in other forms. Hyun \emph{et al.} \cite{hyun2023smile} collected a multimodal dataset, named SMILE, which encompasses video clips including multi-party scenarios. Each video clip is annotated with a reason explaining why the person is laughing. Different than question-answering problem that requires selecting the correct answer from candidates, they generate the laughter reason in a free-form language and measure the performance by calculating the similarity of the generated reason and the annotation. Lai \emph{et al.} \cite{lai2023werewolf} established a multimodal video dataset of people playing the werewolf social deduction game. They propose two problems: (1) predicting which persuasion strategies are used in each utterance and (2) predicting the voting of each player. 

\subsubsection{\textbf{Discussion}} Social situation analysis aims to provide a holistic understanding of the context, norms, and relationships in social interactions. While question answering has been a primary benchmark for evaluating social situation understanding, it has limitations in capturing the complexity of real-world social situations. There also may be some biases between questions and answers \cite{guo2023desiq}.  While alternative approaches like modeling social relationships and utterance strategies provide complementary insights, they still face challenges in addressing potential data bias and incorporating common sense and cultural contexts. The development of more comprehensive and unbiased benchmarks, such as those that incorporate free-form understanding, can be helpful for advancing the field. Moreover, the integration of cultural knowledge and social norms into social situation analysis models can help capture the nuances and variations in social behaviors across different contexts. 

\begin{table*}[t]
\center
\caption{Datasets for understanding multimodal social cues. The table provides attributes of datasets for multimodal emotion analysis, conversation dynamics analysis, and social situation analysis tasks.
}
\vspace{-0.2cm}
\small
\renewcommand{\arraystretch}{0.8}
\renewcommand{\tabcolsep}{0.4mm}
\resizebox{\linewidth}{!}{
\begin{tabular}{c c  c  c  c  c   }
\toprule 
\makecell[l]{Dataset} & {Year} & {Data Types} & {Label Types} & {Data Size}  & {Remarks} \\
\midrule 
\multicolumn{6}{c}{\bf \makecell{Multimodal Emotion Analysis}} \\ \midrule
\makecell[l]{IEMOCAP\cite{busso2008iemocap}} & 2008 & Video, Audio, Text & 9 Discrete Emotions \& 3 Continuous Emotions & 10,000 Videos (12h) & Lab \\
\makecell[l]{SEMAINE \cite{mckeown2011semaine}} & 2012 & Video, Audio, Text & 5 Affective dimensions & 959 Videos (80h) & Lab \\
\makecell[l]{MOSI\cite{zadeh2015micro}} & 2015 & Video, Audio, Text & 7 Sentiment levels & 2,199 Videos (3h) & YouTube \\
\makecell[l]{MOSEI\cite{zadeh2018multimodal}} & 2018 & Video, Audio, Text & 6 Discrete Emotions \& 7 Sentiment levels & 23,453 Videos (66h) & YouTube \\
\makecell[l]{MELD\cite{poria2019meld}} & 2019 & Video, Audio, Text & 8 Discrete Emotions \& 3 Sentiment levels & 1,433 Dialogues (13h) &  TV-series, Multi-party \\

\midrule 
\multicolumn{6}{c}{\bf \makecell{Conversation Dynamics Analysis}} \\ \midrule
\makecell[l]{{SALSA} \cite{alameda2015salsa}} & 2016 & Video, Audio, IR, Accel. & Position \& Pose \& F-formation &  Videos (1h) & Lab, Multi-party \\

\makecell[l]{Triadic Interaction~\cite{joo2019towards}} & 2019 & Video, Depth, Audio, 3D & Speaking status \& Position \& Orientation \& Gesture motion & 180 Videos (3h) &  Lab, Multi-party \\

\makecell[l]{{ConfLab} \cite{raman2022conflab}} & 2022 & Video, Audio, IMU & Pose \&  F-formulation \& Speaking status & 8 Videos (45min) & Lab, Multi-party \\
\makecell[l]{Multi-speaker Conversation~\cite{ryan2023egocentric}} & 2023 & Video, Audio & To whom each person is speaking/listening & Videos (20h) & Lab, Egocentric, Multi-party \\
\makecell[l]{VideoGazeSpeech~\cite{hou2024multi}}  & 2024 & Video, Audio & Gaze target bounding boxes & 29 Videos (23min) & YouTube, Multi-party \\
\makecell[l]{Werewolf Among Us (Exp.)~\cite{lee2024modeling}} & 2024 & Video, Audio, Text & Speaking target \& Pronoun reference \& Mentioned player  & 234 Videos (21h) & YouTube \& Lab, Multi-party \\

\midrule
\multicolumn{6}{c}{\bf \makecell{Social Situation Analysis}} \\ \midrule
\makecell[l]{MovieGraphs \cite{vicol2018moviegraphs}} & 2018 & Video, Audio, Text & Free-text labels about social relationships & 7,637 Videos (94h) &  Movie, Multi-party \\
\makecell[l]{TVQA \cite{lei2018tvqa}} & 2018 & Video, Audio, Text & 152.5K Question \& Answer sets & 21,783 Videos (461h) &  TV-show, Multi-party \\
\makecell[l]{ViSR \cite{liu2019social}} & 2019 & Video, Audio, Text & 8 Social relationships  & 8,000 Videos &  Movie, Multi-party \\
\makecell[l]{Social-IQ \cite{zadeh2019social}} & 2019 & Video, Audio, Text & 7,500 Question \& 52,500 Answer sets & 1,250 Videos (21h) &  YouTube, Multi-party \\
\makecell[l]{LVU \cite{wu2021towards}} & 2021 & Video, Audio, Text & 4 Social relationships  & 30,000 Videos (1,000h) &  Movie, Multi-party \\
\makecell[l]{SMILE \cite{hyun2023smile}} & 2023 & Video, Audio, Text & 887 Explanation sets & 887 Videos (7h) &  TED \& Sitcom, Multi-party \\
\makecell[l]{Werewolf Among Us \cite{lai2023werewolf}} & 2023 & Video, Audio, Text & 6 Persuasion strategies \& Voting outcomes & 252 Videos (22h) &  YouTube \& Lab, Multi-party \\
\bottomrule 
\end{tabular}
}
\vspace{-0.3cm}
\label{table_dataset_multimodal}
\end{table*}

\subsection{Datasets with Multimodal Cues}
Table \ref{table_dataset_multimodal} shows a comprehensive overview, comparing the attributes of datasets with multimodal cues. Further details for each dataset are included in the supplementary material.

\section{Challenges and Future Directions}
As we strive towards machines with advanced social understanding capabilities, there are several key challenges that need to be addressed. In this section, we discuss these challenges and propose future research directions, focusing on the three core capabilities identified earlier: multimodal understanding, multi-party modeling, and belief awareness.

\subsection{Understanding Multimodal Social Cues}

\subsubsection{\textbf{Multimodal Alignment}} Achieving proper alignment between multimodal cues is crucial for effective social understanding. When non-verbal visual cues and verbal language cues are present, they should be aligned not only temporally but also at the person level. This means that when multiple people are present in a visual scene, the visual features of each individual should be matched with their corresponding spoken language, rather than relying solely on global representations. Recent work \cite{lee2024modeling} addressed person-level alignment using visual tracking, but initial manual matching was required. Future research should develop frameworks for automatic person-level alignment, potentially leveraging active speaker detection methods \cite{min2022learning, liao2023light} to dynamically associate individual visuals with spoken utterances.

\subsubsection{\textbf{Multimodal Integration}} The relative importance of different modalities can vary depending on the social context and individuals involved. For instance, visual cues like facial expressions may be more informative than verbal content in some situations, while the opposite may be true in others. Existing multimodal fusion approaches for social reasoning often use fixed weighting schemes \cite{zadeh2019social, lai2023werewolf}. They do not explicitly consider the importance of each modality even when performing intricate feature fusion such as graph-based methods \cite{zadeh2018multimodal, chen2023multivariate}. Future research should investigate sophisticated methods for adaptive multimodal integration based on contextual factors. One possible direction is to compute the feature uncertainty \cite{kendall2017uncertainties} and attention \cite{nagrani2021attention} of each modality and incorporate it into the fusion process to dynamically weight modality importance. In addition, integration through recently emerging general-purpose multimodal foundation models \cite{liu2023llava, achiam2023gpt, team2023gemini} is also an area that has not yet been explored in terms of social understanding. It would be interesting to validate the social capabilities of these models and advance socially intelligent foundation models.

\subsubsection{\textbf{Multimodal Long-range Dependencies}} Social interactions often unfold over extended periods, with long-range dependencies between multimodal cues. For example, a provocative gesture made early in a conversation can affect the overall mood and influence subsequent verbal exchanges. However, most current multimodal social reasoning tasks rely on short videos or a few utterances \cite{ryan2023egocentric, chen2023multivariate, lee2024modeling, zadeh2019social}. There are works that deal with long-term understanding in social contexts \cite{wu2021towards}. Still, their target video length is a few minutes.  Future research should be designed to address the long-term context in multimodal interaction data. Tasks like reasoning about outcomes in social deduction games \cite{lai2023werewolf}, which involve lengthy conversations and dynamics, may serve as effective testbeds. Potential directions include memory augmentation \cite{hazarika2018conversational, lee2021video, wu2022memvit} to efficiently capture long-range dependencies.

\subsubsection{\textbf{Multimodal Understanding Benchmarks.}} While several datasets have been introduced for multimodal social understanding, there remains a need for more comprehensive benchmarks that fully capture the complexity of multimodal interactions. For instance, Social-IQ \cite{zadeh2019social} provides a diverse range of social scenarios. However, there can be some biases between questions and answers \cite{guo2023desiq} due to Q\&A with multiple choice, which allows machines to sometimes solve problems without leveraging multiple modalities. Ego4D \cite{grauman2022ego4d} offers rich egocentric data but is limited in addressing the camera-wearer's own multimodal cues (\emph{e.g.}, their face cannot be observed). MovieGraphs \cite{vicol2018moviegraphs} provides detailed graph-based annotations of social situations but the scenes are mainly staged and may not fully represent real-world, natural interactions. Werewolf-Among-Us \cite{lai2023werewolf, lee2024modeling} captures complex group dynamics including multimodal social cues but are confined to specific game contexts. Future efforts should focus on creating large-scale datasets with diverse social interactions and fine-grained multimodal annotations across different modalities (\emph{e.g.}, text, audio, video). Moreover, the datasets should include challenging reasoning tasks that require the fine-grained multimodal alignment and integration of multimodal information over extended periods.

\subsection{Understanding Multi-party Dynamics}

\subsubsection{\textbf{One-to-many Dynamics}} Real social scenarios often involve individuals engaging in one-to-many interactions, such as a speaker addressing multiple listeners simultaneously. However, existing works mainly focus on one-to-one interaction dynamics \cite{jia2024audio, lee2024modeling} even in multi-party environments. Modeling one-to-many dynamics is crucial for capturing the full complexity of social interactions. Future research should develop techniques to represent and reason about the dependencies between an individual and multiple interaction partners. This may involve creating new benchmarks specifically designed for one-to-many scenarios or extending graph neural networks \cite{jia2024audio, chen2023multivariate} to handle multiple interaction targets.

\subsubsection{\textbf{Group-level Dynamics}} Social dynamics can emerge at the group level, where multiple individuals or subgroups interact with each other, such as in team meetings or debates. Modeling group-level dynamics requires capturing the complex web of relationships and influences among group members, as well as the collective behaviors and properties of the group as a whole. Future work should explore techniques for representing and analyzing group structures, such as hypergraph neural networks \cite{feng2019hypergraph} or hierarchical graph models \cite{xing2021learning}, that can capture interactions at multiple scales. Social deduction games with team dynamics can serve as good testbeds \cite{lai2023werewolf}.

\subsubsection{\textbf{Interaction History Modeling}} Current conversation dynamics is influenced by their interaction history between participants. For instance, knowing who spoke to whom, who looked at whom, and who reacted to whom in previous interactions can provide important clues about ongoing conversation dynamics and interpersonal relationships. Future research can explore methods for encoding and integrating multimodal interaction history into multi-party models. This may involve extending temporal graph embeddings \cite{rossi2020temporal, ghosal2019dialoguegcn} to incorporate multiple modalities in encoding interaction history.

\subsubsection{\textbf{Multi-party Understanding Benchmarks}} While some datasets have been introduced for multi-party interaction analysis, there are still gaps in the comprehensiveness and diversity of these benchmarks, which is corroborated by findings from \cite{li2024social}. The Social-IQ dataset \cite{zadeh2019social}, although containing multi-party social videos, primarily focuses on question-answering tasks and lacks fine-grained annotations for modeling the relationships and interactions between participants. Werewolf-Among-Us dataset \cite{lai2023werewolf, lee2024modeling}, while capturing complex and fine-grained multi-party dynamics, is limited to specific game scenarios and may not generalize to other social contexts. An ideal multi-party interaction benchmark should cover a wide range of real-world multi-party scenarios, such as group discussions, team meetings, and social gatherings, with varying numbers of participants and group structures. The dataset should include rich annotations for tasks like identifying the referents in utterances, recognizing the relationships between participants, and predicting the group dynamics such as alliances or conflicts. Moreover, the benchmark should include longitudinal interaction data to facilitate the study of interaction history and group dynamics over time.

\subsection{Understanding Beliefs}

\subsubsection{\textbf{Cultural Awareness}} Cultural background significantly shapes beliefs, norms, and expectations, influencing their behavior and social interactions. However, current multimodal interaction understanding approaches often overlook cultural aspects. Future research should prioritize cultural awareness in social interaction models by incorporating cultural knowledge and adapting to cultural differences. Future research may extend techniques like culture-aware language inference \cite{huang2023culturally} in multimodal aspects including non-verbal social cues.

\subsubsection{\textbf{Relationship Awareness}} The relationship between interacting individuals, such as their social roles, power dynamics, or emotional connections, affects their shared beliefs and plays a significant role in shaping interaction dynamics. However, current multimodal approaches often treat all interactions uniformly without considering underlying relationships. Even though there exist works that deals with social relationship inference \cite{kukleva2020learning, wu2021linking, wang2023shifted}, most of them do not effectively incorporate predicted social relationships to interpret further social interactions. Future work should develop techniques integrating relationship knowledge between interacting parties in social reasoning. This may involve extending methods like relational graph neural networks \cite{schlichtkrull2018modeling, goel2019end} that can learn relationship-aware representations. Moreover, future research also can explore ways to incorporate external knowledge about relationships into social understanding.

\subsubsection{\textbf{Individual Belief Awareness}} Understanding an individual's beliefs, intentions, and mental states is crucial for interpreting their behaviors in social interactions. Current methods for individual belief modeling primarily rely on language-based cues. Future research should focus on developing techniques that integrate both verbal and non-verbal dynamics to infer individual beliefs and mental states from multimodal interaction data. This may involve approaches like theory of mind modeling \cite{rabinowitz2018machine}, which aims to simulate cognitive processes of perspective-taking and mental state attribution, or inverse reinforcement learning \cite{jara2019theory} methods that can infer an individual's goals and preferences from their observed behavior.

\subsubsection{\textbf{Belief-aware Benchmarks}} Existing multimodal interaction datasets often lack the diversity and granularity of annotations needed to effectively train and evaluate models that incorporate individual beliefs, social relationships, and cultural contexts. For example, while the Triadic Belief dataset \cite{fan2021learning} provides some annotations related to beliefs, it is limited to a narrow set of belief types in specific lab-based interaction scenarios. Similarly, datasets like Social-IQ \cite{zadeh2019social} and Werewolf-Among-Us \cite{lai2023werewolf, lee2024modeling}, although valuable for studying social interactions, do not provide detailed annotations about the beliefs, intentions, and mental states of individual participants, nor do they explicitly capture information about social relationships or cultural backgrounds that influence those beliefs. Ideal belief-aware benchmarks should encompass a diverse range of real-world social interaction scenarios with varying cultural contexts, social relationships, and individual characteristics. Crucially, it should include fine-grained annotations for various aspects of beliefs at both the individual and group levels. This would include labels for intentions, goals, and mental states of each participant throughout the interaction, as well as annotations for the relationships between participants and their cultural backgrounds.

\section{Conclusion}
In this survey, we provide a comprehensive overview of the current state of research in understanding social interactions through artificial intelligence. We examine the landscape of existing work across three key modalities: verbal cues, non-verbal cues, and multimodal approaches. Our analysis highlights the importance of three core capabilities for effective social understanding: multimodal integration, multi-party modeling, and belief awareness. Looking ahead, we outline several critical challenges and promising future directions in each of these areas. These include improving multimodal alignment and integration, modeling complex group dynamics, and incorporating cultural and individual belief awareness into social interaction models. As the field progresses, addressing these challenges will be crucial for developing AI systems capable of truly understanding and seamlessly participating in human social interactions. By advancing along these research directions, we can move closer to the goal of creating socially intelligent AI that can effectively interpret, reason about, and engage in the multifaceted nature of human social interactions.

\vspace{-0.3cm}
\ifCLASSOPTIONcompsoc
  \section*{Acknowledgments}
  
\else
  \section*{Acknowledgment}
\fi

This work was supported in part by NSF CNS-2308994.

\ifCLASSOPTIONcaptionsoff
  \newpage
\fi

\bibliographystyle{IEEEtran}
\bibliography{ref_file}
\label{sec:refs}

\clearpage



\section*{Supplementary material}

## Supplementary Material

# 1 Datasets with Verbal Cues

## 1.1 Dialogue Act Analysis Datasets

Different datasets on dialogue act analysis usually focus on different domains such as train booking, restaurant search, counseling and so on. They are labeled with domain-specific dialogue acts, and systems are evaluated using the accuracy in correctly classifying the dialogue act labels.

**SwitchboardDA** *Switchboard Dialog Act Corpus* [1] contains 1,155 five-minute dialogues between two speakers on telephone, resulting in a total of 221,616 utterances. It contains speaker-level, utterance-level and conversation-level of information and is annotated with approximately 60 different speech act tags.

**MRDA Corpus** *ICSI Meeting Recorder Dialog Act Corpus* [2] consists of 72 hours of speech with over 180,000 manually annotated dialog act tags. The dialogues happen in meetings with an average of six speakers. The corpus is annotated with dialogue act segment boundaries, dialogue acts, and adjacency pairs.

**AMI Corpus** [3] contains 100 hours of video recordings of scenario-driven meetings. Participants were assigned the roles of employees in a company who were discussing the development of a new product. The dialogues consist of four phases: project kick-off, functional design, conceptual design, and detailed design. The data comes with annotations in terms of transcriptions, dialogue acts and summaries.

**DSTC** *Dialog State Tracking Challenge* dataset [4] provides a collection of 1,500 human-computer dialogues with a set of 11 evaluation metrics. It uses the conversation data from the public deployment of several systems. Each dialogue is annotated with dialogue states that represent user's goals and intentions.

**DSTC2** *The second Dialog State Tracking Challenge* [5] introduces some new features. It introduces a new domain

of restaurant search. The users' goals are also allowed to change during the dialogues. Moreover, it uses a richer representation of dialog states which covers the search method and information the user would like the system to read out. It contains 3235 dialogues in total.

**DSTC3** *The third Dialog State Tracking Challenge* [6] contains 2,275 dialogues from paid crowdworkers. It introduces a new domain on tourist information. Also, there are new entity types and values in the test data which do not appear in the train data, posing greater challenges in terms of generalization capability of models.

**HOPE** [7] contains around 12,900 utterances from counseling transcripts from publicly available Youtube videos. The texts are annotated with 12 different dialogue act labels. They can be grouped into speaker initiative labels, speaker responsive labels and general labels.

**IMCS-21** [8] is a dialogue dataset on medical consultation consisting of 4,116 annotated dialogues with 164,731 utterances. The data comes from online professional medical consulting community. There are two broad dialogue act categories, which are inform and request. It is also annotated with more fine-grained categories such as basic information, symptom, etiology, existing exam and treatment.

## 1.2 Dialogue Emotion Analysis Datasets

Researchers have created various dialogue datasets for emotion analysis. The dialogues can be dyadic and multi-party. The number of emotion classes in each dataset can vary depending on different annotation schemes. Systems are evaluated using the accuracy score and F1 score in classifying the emotions.

**DailyDialogue** [9] consists of 13,118 human-written high-quality dialogues. The dialogues are in the multi-turn format with rich contextual information, where most previous dialogue datasets did not have such format. The dialogues are daily conversations on a variety of topics across multiple

TABLE 1: Datasets for understanding verbal social cues. The table provides attributes of datasets for dialogue act analysis, dialogue emotion analysis, and common sense reasoning tasks.

| Dataset | Year | Data Types | Label Types | Data Size | Remarks |
|---|---|---|---|---|---|
| **Dialogue Act Analysis** | | | | | |
| SwitchboardDA [1] | 2000 | Audio, Text | Around 60 speech tags | 1,155 Conversations | Telephone call |
| MRDA Corpus [2] | 2004 | Audio | 39 Specific dialogue act tags | 72 Hour meetings | Group meeting, Multi-party |
| AMI Corpus [3] | 2005 | Video | 15 Dialogue act tags | 100 Hour meetings | Group meeting, Multi-party |
| DSTC [4] | 2013 | Audio, Text | Varying number of dialogue states | 1,500 Conversations | Telephone call |
| DSTC2 [5] | 2014 | Audio, Text | Varying number of dialogue states | 3,235 Conversations | Telephone call |
| DSTC3 [6] | 2014 | Audio, Text | Varying number of dialogue states | 2,275 Conversations | Telephone call |
| HOPE [7] | 2022 | Text | 12 Dialogue acts | 12,900 Utterances | Counselling session |
| IMCS-21 [8] | 2023 | Text | 8 Specific dialogue act tags | 4,116 Dialogues | Medical consultation |
| **Dialogue Emotion Analysis** | | | | | |
| DailyDialogue [9] | 2017 | Text | 6 Emotion classes | 13,118 Dialogues | Daily dialogue, Multi-turn |
| EmotionLines [10] | 2018 | Text | 7 Emotion classes | 2,000 Dialogues | TV-show & Messenger, Multi-party |
| EmoContext [11] | 2019 | Text | 4 Emotion classes | 38,424 Dialogues | User-Agent interaction |
| EmoWOZ [12] | 2022 | Text | 7 Emotion classes | 11,434 Dialogues | User-Agent interaction, Multi-turn |
| **Common Sense Reasoning** | | | | | |
| WS Challenge [13] | 2012 | Text | 2 Disambiguation options to choose | 273 Sentences | Coreference resolution |
| SNLI [14] | 2015 | Text | 3 Types of logical connections | 570,000 Pairs | Text entailment |
| Story Cloze Test [15] | 2016 | Text | 2 Endings and label indicating the correct ending | 3742 Stories | Stories ending inference |
| JOCI [16] | 2017 | Text | 5 Plausibility labels | 39,093 Pairs | Psychological inference |
| Story Commonsense [17] | 2018 | Text | 5 Motivation classes & 8 emotion classes | 15,000 Stories | Psychological inference |
| Event2Mind [18] | 2018 | Text | People's possible intents & reactions | 24,716 Events | Psychological inference |
| CommonsenseQA [19] | 2019 | Text | MCQ questions | 12,247 Questions | Question Answering |
| SocialIQA [20] | 2019 | Text | MCQ questions | 38,000 Questions | Question Answering |

domains. They are annotated with communication functions and six categories of emotions.

**EmotionLines** [10] contains 2000 multi-party dialogues from Friends TV scripts, resulting in a total of 29,245 utterances. They are labeled with six Ekman's basic emotions and one neutral emotion by five Amazon MTurkers.

**EmoContext** [11] was published as a task in SemEval-2019. It presents the dialogue context with two previous turns and contains annotation of four emotion classes (Happy, Sad, Angry and Others). There is a total of 38,424 dialogues.

**EmoWOZ** [12] is built based on previous established multi-domain task-oriented dialogue dataset MultiWOZ [21]. On top of MultiWOZ, the authors collected human-machine dialogues in the same domains to ensure a better coverage of different emotion types. In total, there are more than 11,000 dialogues and 83,000 emotion annotations.

### 1.3 Common Sense Reasoning

Common sense reasoning benchmarks usually contain a context such as event and premise, provide possible options for disambiguation of coreference and possible inference on people's mental states and plausible following events. They evaluate systems' inference capability using classification accuracy in terms of accuracy scores and F1 scores.

**WS Challenge** *Winograd Schema Challenge* [13] is built by selecting one sentence from each sentence pair created under Winograd schema. Sentences in the same pair are different from each other in just one or two words, which lead to different referential resolutions in the sentences. Models need common sense reasoning skills to achieve human-level accuracy in terms of resolving the ambiguities correctly.

**SNLI** Stanford Natural Language Inference [14] holds a collection of 570,000 sentence pairs consisting of a premise and a hypothesis, which are annotated with the logical connection in terms of entailment, contradiction and neutral. Each pair is annotated by five different crowdworkers. Understanding the semantic relationship between two sentences in a pair require the common sense reasoning ability.

**Story Cloze Test** [15] contains stories randomly sampled from ROCStories Corpus. The crowdworkers will be presented with the first four sentences of each story and annotate the story with a right ending and a wrong ending based on common sense. The annotated stories will be verified by three crowdworkers for quality control, resulting in a final set of 3,742 test cases.

**JOCI** JHU Ordinal Commonsense Inference [16] contains 39,093 sentence pairs annotated with an ordinal scale of plausibility. The authors sampled contexts from SNLI and ROCStories and generated possible inference candidates for those contexts using general world knowledge or neural methods. After that, the sentence pairs were annotated with level of plausibility via crowdsourcing.

**Story Commonsense** [17] holds a collection of 15,000 annotated stories under a carefully designed annotation pipeline. The stories have annotations on entities, actions, affects, motivations and emotion reactions. It can benchmark models' common sense reasoning performance in classifying motivations and emotion reactions and providing explanations.

**Event2Mind** [18] consists of 24,716 phrases on a wide variety of daily events. The events are extracted from ROCStories, Google Syntactic N-grams and Spinn3r corpus. They are annotated with possible intents and reactions of the events' participants via crowdsourcing.

**CommonsenseQA** [19] contains 12,247 questions that require the common sense of real world knowledge. The dataset contains data in MCQ formats. The instances are designed in a challenging way where similar concepts in ConceptNet are extracted and semantically difficult questions are designed by crowdworkers.

**SocialIQA** [20] is on common sense reasoning for social situations. There are 38,000 MCQs in the dataset about social situations. The questions and answers are collected from crowdworkers. To address the potential problem of annotation artifacts during crowdsourcing, the authors combined manually created negative answers and adversarial question-switched answers.

## 2 DATASETS WITH NON-VERBAL CUES

### 2.1 Gesture Datasets

The majority of gesture recognition datasets are based on video data. Recent advancements in research underscore a new trend by increasing the distance between people and the capturing devices to increase models' robustness [27], [29]. In this subsection, we introduce 9 public benchmarks and datasets crucial for gesture recognition research. The evaluation metrics used to assess model efficacy on these benchmarks include some classification metrics like Accuracy, Mean Average Precision (mAP), and the F1-score. For tasks related to continuous gesture recognition [22], [27], the Jaccard Index or Mean Jaccard Index are employed to provide a thorough evaluation of a model's ability to accurately segment continuous gestures. The Jaccard Index measures the proportion of overlap between the actual gesture segment and the model's predicted segment over the total number of frames, providing a quantitative measure of accuracy for gesture recognition.

**ChaLearn IsoGD** Both ChaLearn IsoGD and ChaLearn ConGD [22] trace their origins to the CGD 2011 [52]. The ChaLearn IsoGD dataset is a large-scale dataset dedicated to isolated hand gestures. It encompasses 47,933 RGB-D gesture videos. Each RGB-D video uniquely represents a single gesture, with a total of 249 gesture labels performed by 21 different individuals. Each gesture category is represented by over 200 RGB and depth videos.

**ChaLearn ConGD** [22], developed alongside the ChaLearn IsoGD dataset, focuses on continuous gesture recognition. It comprises 22,535 RGB-D videos, capturing a total of 47,933 gestures. Unlike isolated gestures, each RGB-D video in the ChaLearn ConGD dataset may encompass one or several gestures, spanning across 249 distinct gesture labels executed by 21 different participants.

**EgoGesture** [23] is a large-scale, multimodal collection designed for egocentric perspective. It encompasses 2,081 RGB-D videos from 50 distinct subjects, containing 24,161 gesture instances and 2,953,224 frames. The dataset features 83 types of static or dynamic gestures, tailored for interactions with wearable devices. These videos were collected from 6 varied indoor and outdoor settings.

**Jester** [24] features 148,092 video clips annotated with 27 gesture categories, showcasing individuals executing predefined gestures in front of web cameras, including actions

TABLE 2: Datasets for understanding non-verbal social cues. The table provides attributes of datasets for gesture and body language analysis, gaze and attention analysis, and facial expression tasks.

| Dataset | Year | Data Types | Label Types | Data Size | Remarks |
|---|---|---|---|---|---|
| **Gesture and Body Language Analysis** | | | | | |
| ChaLearn IsoGD [22] | 2016 | Video | 249 Gesture classes | 47,933 Videos | Lab, Isolated |
| ChaLearn ConGD [22] | 2016 | Video | 249 Gesture classes | 22,535 Videos | Lab, Continuous |
| EgoGesture [23] | 2018 | Video | 83 Gesture classes | 2,081 Videos (27h) | Lab, Egocentric, Isolated |
| Jester [24] | 2019 | Video | 27 Gesture classes | 148,092 Videos (123h) | Lab, Webcam, Isolated |
| IPN Hand [25] | 2020 | Video | 13 Gesture classes | 200 Videos (7h) | Lab, Isolated |
| iMiGUE [26] | 2021 | Video | 32 Gesture classes | 359 Videos (35h) | Web, Interview, Isolated |
| LD-ConGR [27] | 2022 | Video | 10 Gesture classes | 542 Videos | Lab, Continuous |
| SMG [28] | 2023 | Video | 17 Gesture classes | 40 Videos (8h) | Lab, Isolated & Continuous |
| HaGRID [29] | 2024 | Image | 18 Gesture classes | 552,992 Images | Lab, Isolated |
| **Gaze and Attention Analysis** | | | | | |
| GazeFollow [30] | 2015 | Image | Gaze point | 122,143 Images | Web, Multi-party |
| VideoGaze [31] | 2017 | Video | Gaze point & in/out of frame | 140 Videos (166,721 frames) | Movie, Multi-party |
| DLGaze [32] | 2018 | Video | Gaze point | 86 Videos (53min) | Lab, Multi-party |
| VideoCoAtt [33] | 2018 | Video | Shared attention bounding boxes | 380 Videos (5.5h) | YouTube, Multi-party |
| EGTEA Gaze+ [34] | 2018 | Video | Eye tracking | 86 Videos (28h) | Lab |
| VACATION [35] | 2019 | Video | Gazed object boxes & Gaze classes (6 atomic/5 event) | 300 Videos (1h) | YouTube, Multi-party |
| UCO-LAEO [36] | 2019 | Video | Mutual for each pair of people | 129 Videos (10min) | TV-show, Multi-party |
| AVA-LAEO [36] | 2019 | Video | Mutual for each pair of people | 298 Videos (13h) | Movie, Multi-party |
| VideoAttentionTarget [37] | 2020 | Video | Gaze point + in/out of frame | 606 Videos (71,666 frames) | YouTube, Multi-party |
| Triadic Belief [38] | 2021 | Video | Eye tracking, pointing, events (3 types), beliefs (4 types) | 88 Videos (72m) | Lab, Multi-party |
| Ego4D [38] | 2022 | Video | Eye tracking | 27 Videos (31h) | Lab, Multi-party |
| ChildPlay [39] | 2023 | Video | Gaze point + in/out of frame | 401 Videos (1h) | YouTube, Multi-party |
| **Facial Expression Analysis** | | | | | |
| MMI [40], [41] | 2005 | Image, Video | 7 Expression classes | 740 Images, 2,900 Videos | Lab, Static & Dynamic |
| CK+ [42] | 2010 | Video | 8 Expression classes | 593 Videos | Lab, Dynamic |
| FER-2013 [43] | 2013 | Image | 7 Expression classes | 35,887 Images | Web, Static |
| SFEW 2.0 [44] | 2015 | Image | 7 Expression classes | 1,766 Images | Movie, Static |
| EmotioNet [45] | 2016 | Image | 23 Expression classes | 1,000,000 Images | Web, Static |
| AFEW 7.0 [46] | 2017 | Video | 7 Expression classes | 1,809 Videos | Movie, Dynamic |
| RAF-DB [47], [48] | 2017 | Image | 7 Expressions & 12 Compound expressions | 29,672 Images | Web, Static |
| AffectNet [49] | 2017 | Image | 7 Expression classes | 450,000 Images | Web, Static |
| ExpW [50] | 2018 | Image | 7 Expression classes | 91,793 Images | Web, Static |
| Face MPMI [51] | 2019 | Video | 7 Expression classes | 37,856 Videos | Lab, Dynamic |

like two-finger swipes down, left or right swipes, and finger taps. The inclusion of two "no gesture" categories enhances the model's ability to differentiate between specific gestures and incidental hand movements.

**IPN Hand** [25] comprises over 4,000 instances of hand gestures across 800,000 frames from 50 participants. It encompasses 13 types of static and dynamic gestures designed for interactions with non-touchscreen interfaces. The IPN Hand features the highest rate of intra-class variation and also offers real-time optical flow and hand segmentation results.

**iMiGUE** [26] is the first public dataset for emotional micro-gestures, which are involuntary actions motivated by underlying emotions. The dataset is annotated on two levels: the micro-gesture categories were annotated on video clip-level, and the emotion categories were labeled on video-level. The iMiGUE organizes 32 distinct categories for 359 videos from post-match press conferences.

**LD-ConGR** [27] features a total of 542 videos and 44,887 instances of hand gestures. These videos were collected from 30 participants across five distinct environments, using a Kinect V4 camera to capture footage from a third-person perspective. Each video comprises both a color stream and a depth stream, which are synchronized.

**SMG** [28], designed for micro-gesture recognition, consists of 40 video sequences (8 hours total) from 40 participants. It includes RGB, depth, silhouette videos, and skeleton data. The dataset contains 3712 pre-segmented and labeled

micro-gesture clips across 16 classes and non-micro-gesture classes. It also includes 71 instances of relaxed and stressed emotional states, each lasting 2-5 minutes.

**HaGRID** [29] is a large-scale, high-resolution dataset for static gesture classification or detection. It contains 552,992 FullHD RGB images across 18 hand gesture classes. The dataset was primarily collected indoors, with some images taken under extreme lighting conditions. Participants performed gestures at distances ranging from 0.5 to 4 meters from the camera for diverse spatial representations.

### 2.2 Gaze Datasets

Various gaze analysis datasets address the tasks of predicting spatial gaze targets, where predictions are evaluated against ground truth gaze points, and recognizing multi-person gaze behaviors, where predictions are evaluated at the frame or instance level.

**GazeFollow** [30] consists of 122,143 internet images annotated with head bounding boxes and gaze points. The train set has one gaze annotation per instance, while the test set has $\approx 10$ unique annotations per instance. In total, there are 130,339 instances (head bounding box with paired gaze target).

**VideoGaze** [31] addresses the unique case of identifying gaze targets in following video frames, where the camera often pans from the person to the gaze target. Each annotation consists of 6 frames, where the head bounding box is provided for the first frame and the gaze target locations are

annotated for the following 5 frames (including indication of if the target is out of frame). It is sourced from 140 movies.

**DLGaze** [32] consists of 86 videos manually collected in 4 unique locations by 16 people. The participants themselves then annotated the location of their gaze targets. In total there are 95,000 frames. 4 frames are annotated per second, so there are $\approx 6,333$ instances.

**VideoCoAttention** [33] is a video dataset for identifying shared attention, which occurs when two or more people are looking at the same target. Shared attention instances are annotated as a set of head bounding boxes for the people involved and a bounding box for the shared target of gaze. In total, there are 138,203 instances of shared attention annotated across 492,1000 frames from 380 videos of TV shows / movies.

**EGTEA Gaze+** [34] is an egocentric dataset consisting of 28 hours of people cooking in a kitchen environment. It is collecting with wearable eye tracking glasses, providing ground truth for egocentric gaze estimation.

**VACATION** [35] is a video dataset for recognizing a broader set of multi-party communicative gaze behaviors. It defines 6 atomic-level behaviors: single, mutual, avert, refer, follow, and share, as well as 5 event-level gaze behaviors: non-communicative, mutual gaze, gaze eversion, gaze following, and joint attention. The dataset contains 300 video sequences from movies and TV shows, with a total of 96,993 frames.

**UCO-LAEO & AVA-LAEO** [36] are benchmarks for identifying mutual gaze, or "looking at each other". They contain a total of 208,688 instances comprised of head bounding boxes for two individuals and a binary label indicating whether they are looking at each other. UCO-LAEO has $\approx 18,000$ frames from 4 TV shows, while AVA-LAEO includes $\approx 1.4M$ frames from 298 movies, sourced from the AVA dataset [53].

**VideoAttentionTarget** [37] is comprised of 606 video segments from 50 TV shows or movies. Annotations consist of head bounding box tracks with an annotated gaze point. Additionally, VideoAttentionTarget labels if the gaze point is in or out of frame and includes this as an additional benchmark task. In total, there are 71,666 frames and 164,541 instances.

**Triadic Belief** [54] contains 88 videos (109,331 frames) of social interactions captured from both third-person and first-person views. It's annotated with gaze, gestures, communication events, and belief dynamics.

**Ego4D** [38] contains 31 hours of egocentric video collected with wearable eye tracking glasses, which is used as a benchmark for egocentric gaze estimation. This eye tracking subset was collected in multi-person social settings.

**ChildPlay** [39] targets on the specific domain of children interacting with adults and other children, sourcing videos from Youtube. Annotations follow the same scheme as VideoAttentionTarget, consisting of the head bounding box, annotated gaze point, and in vs. out of frame label.

## 2.3 Facial Expression Datasets

Having sufficiently large and labelled dataset is essential to ensure the optimal learning of deep facial expression recognition (FER) methods. This dataset should encompass a diverse range of variations in both the surrounding environment and facial structures. In this section, we delve into the discussion of public benchmark FER databases, commonly utilized in the researched papers, which cover fundamental emotions, which are classified into seven categories: happy, anger, fear, disgust, sad, surprise, neutral. The performance is evaluated using several metrics, including precision, accuracy, recall, specificity, and F1-score.

**MMI** [40], [41] is a laboratory-controlled dataset containing 326 sequences from 32 subjects. It includes 213 sequences annotated with six basic expressions, with 205 sequences captured in a frontal view. MMI sequences start from a neutral expression, reach a peak, and return to a neutral state. The dataset presents challenges due to inter-personal variations, non-uniform expressions, and subjects wearing accessories.

**CK+** [42] is a facial expression dataset captured in a controlled lab environment. It contains 593 video sequences of 123 subjects, with durations ranging from 10 to 60 frames, showing the transition from neutral to peak emotional expressions. 327 sequences from 118 subjects are annotated with seven basic emotion labels using the Facial Action Coding System (FACS).

**FER-2013** [43] consists of 35,887 grayscale images categorized into seven different facial expressions. The images are relatively low-resolution (48x48 pixels) and depict diverse individuals across different age groups and ethnicities, consisting of both posed and unposed headshots.

**SFEW** *The Static Facial Expressions in the Wild* [44] dataset was generated by selecting static frames from the AFEW database through the computation of key frames based on facial point clustering. Primarily used as benchmark data for the SReco sub-challenge in EmotiW 2015, it consists of 1,766 samples. Each image is labeled with one of seven expression categories

**EmotioNet** EmotioNet [45] is an extensive dataset that comprises one million facial images sourced from the Web. 950,000 umages were annotated using an automatic action unit detection model whereas the remaining 25,000 images underwent manual annotation, specifically identifying 11 action units (AUs).

**AFEW** *The Acted Facial Expressions in the Wild* [46] contains video clips from movies with spontaneous expressions, diverse head poses, occlusions, and varying illuminations. It serves as a temporal and multimodal database with audio and video data. The dataset includes samples labeled with seven expressions from reality TV shows. For EmotiW 2017, AFEW 7.0 consists of 1,809 samples.

**RAF-DB** *Real-world Affective Faces Database* [47], [48] is a comprehensive dataset containing 29,672 facial images with basic or compound expression labels, annotated by 40 independent taggers. The dataset features diverse subjects, varied head poses, lighting conditions, occlusions, and post-processing effects, making it a rich resource for studying facial expression recognition in real-world scenarios.

**AffectNet** AffectNet [49] comprises over one million images sourced from Web by querying different search engines with keywords related to facial expressions. 450,000 images have manually annotated for eight expressions, and it is the largest dataset providing images with facial expressions for both categorical and dimensional emotional models.

TABLE 3: Datasets for understanding multimodal social cues. The table provides attributes of datasets for multimodal emotion analysis, conversation dynamics analysis, and social situation analysis tasks.

| Dataset | Year | Data Types | Label Types | Data Size | Remarks |
|---|---|---|---|---|---|
| **Multimodal Emotion Analysis** | | | | | |
| IEMOCAP [55] | 2008 | Video, Audio, Text | 9 Discrete Emotions & 3 Continuous Emotions | 10,000 Videos (12h) | Lab |
| SEMAINE [56] | 2012 | Video, Audio, Text | 5 Affective dimensions | 959 Videos (80h) | Lab |
| MOSI [57] | 2015 | Video, Audio, Text | 7 Sentiment levels | 2,199 Videos (3h) | YouTube |
| MOSEI [58] | 2018 | Video, Audio, Text | 6 Discrete Emotions & 7 Sentiment levels | 23,453 Videos (66h) | YouTube |
| MELD [59] | 2019 | Video, Audio, Text | 8 Discrete Emotions & 3 Sentiment levels | 1,433 Dialogues (13h) | TV-series, Multi-party |
| **Conversation Dynamics Analysis** | | | | | |
| SALSA [60] | 2016 | Video, Audio, IR, Accel. | Position & Pose & F-formation | Videos (1h) | Lab, Multi-party |
| Triadic Interaction [61] | 2019 | Video, Depth, Audio, 3D | Speaking status & Position & Orientation & Gesture motion | 180 Videos (3h) | Lab, Multi-party |
| ConfLab [62] | 2022 | Video, Audio, IMU | Pose & F-formulation & Speaking status | 8 Videos (45min) | Lab, Multi-party |
| Multi-speaker Conversation [63] | 2023 | Video, Audio | To whom each person is speaking/listening | Videos (20h) | Lab, Egocentric, Multi-party |
| VideoGazeSpeech [64] | 2024 | Video, Audio | Gaze target bounding boxes | 29 Videos (23min) | YouTube, Multi-party |
| Werewolf Among Us (Exp.) [65] | 2024 | Video, Audio, Text | Speaking target & Pronoun reference & Mentioned player | 234 Videos (21h) | YouTube & Lab, Multi-party |
| **Social Situation Analysis** | | | | | |
| MovieGraphs [66] | 2018 | Video, Audio, Text | Free-text labels about social relationships | 7,637 Videos (94h) | Movie, Multi-party |
| TVQA [67] | 2018 | Video, Audio, Text | 152.5K Question & Answer sets | 21,783 Videos (461h) | TV-show, Multi-party |
| ViSR [68] | 2019 | Video, Audio, Text | 8 Social relationships | 8,000 Videos | Movie, Multi-party |
| Social-IQ [69] | 2019 | Video, Audio, Text | 7,500 Question & 52,500 Answer sets | 1,250 Videos (21h) | YouTube, Multi-party |
| LVU [70] | 2021 | Video, Audio, Text | 4 Social relationships | 30,000 Videos (1,000h) | Movie, Multi-party |
| SMILE [71] | 2023 | Video, Audio, Text | 887 Explanation sets | 887 Videos (7h) | TED & Sitcom, Multi-party |
| Werewolf Among Us [72] | 2023 | Video, Audio, Text | 6 Persuasion strategies & Voting outcomes | 252 Videos (22h) | YouTube & Lab, Multi-party |

**ExpW** The Expression in-the-Wild (ExpW) [50] dataset is a collection of facial images specifically labeled for expressions. The dataset includes 91,793 faces, each manually annotated with one of seven basic expression categories. Notably, ExpW stands out for its extensive size and diverse facial variations, surpassing many existing databases in terms of the breadth of facial expressions captured.

**Face MPMI** The Face Multi-Pose Multi-Illumination (Face MPMI) [51] dataset contains dynamic facial expressions from 104 subjects, captured under 13 pose variations, 4 illumination conditions (room, bright, left, and right), and with/without eyeglasses. This comprehensive dataset enables evaluation under diverse challenging conditions.

## 3 DATASETS WITH MULTIMODAL CUES

### 3.1 Multimodal Emotion Datasets

As the field of multimodal emotion analysis progresses, a variety of sentiment analysis and emotion recognition datasets have been introduced to facilitate research and development in this area. All of these datasets consist of annotated data that offer video, audio, and transcribed texts as available modalities, while some of them also include labels for facial expressions or gestures. Classification categories vary across datasets; however, Ekman's six universal emotions [73] (Joy, Sadness, Fear, Anger, Surprise, and Disgust) serve an important role in category design. Given the subjective and biased nature of emotion judgment, it is common practice to preset discrete emotion categories and merge annotation results from multiple annotators using a majority voting scheme. Additionally, Mean Absolute Error (MAE) and F1-Score are commonly used as evaluation metrics. In this section, we will provide a brief overview of four widely used multimodal datasets for sentiment/emotion analysis.

**IEMOCAP** [55] contains 12 hours of audio-visual data from 5 dyadic sessions acted by 10 actors. The data includes motion capture of facial expressions, dialog transcriptions, video, and audio recordings. The emotions are expanded from six universal emotions [73] to include fear, excitement,

and surprise. Primitive attributes such as valence, activation, and dominance are also included.

**SEMAINE** The SEMAINE dataset [56] contains 959 conversations with Sensitive Artificial Listener characters. The data is audiovisual recordings of 150 participants with text transcriptions. Each clip is annotated by six to eight raters among 27 classes along 5 emotional dimensions.

**MOSI** [57] consists of 93 opinion videos from YouTube, featuring 89 distinct speakers (41 females and 48 males). The videos are segmented into 2,199 subjective and 1,503 objective opinion segments. Only the subjective segments are annotated with with sentiment annotations and intensity on a linear scale from +3 (strongly positive) to -3 (strongly negative) by two trained annotators.

**MOSEI** [58] is an extension of MOSI, with 23,453 video segments featuring 1,000 speakers and 250 topics. It adopts MOSI's sentiment intensity labeling and adds emotion labels. Apart from incorporating the six Ekman emotions [73], it establishes a linear scale to indicate the possible presence of these emotions. Annotations are provided by three Amazon Mechanical Turk annotators.

**MELD** [59] is a multimodal extension of Emotionlines [10], sourcing data from the TV series *"Friends"*. It consists of 1,433 dialogues (13,000 utterances), including multi-party conversations. Ekman's six emotions are expanded with Neutral and Non-Neutral labels. Additional sentiment classes (positive, negative, neutral) are utilized to further annotate complex emotions such as surprise.

### 3.2 Conversation Dynamics Analysis Datasets

The dataset for conversation dynamics analysis may have a large variation since the dynamic nature of conversations can be observed in different aspects. Typically, we consider several datasets of social scenarios that include RGB frames and human's speech. These datasets may vary in the tasks (such as classification, localization, etc.) but they are all used to study how the speech impacts human's intention and attention as well as its dynamic nature.

**SALSA** [60] dataset contains 1-hour recordings of two unconstrained social scenes: a poster session and a cocktail

party. Videos are captured from 4 synchronized viewpoints using a surveillance system, resulting in distant participants and noisy audio. The dataset includes audio, accelerometer data, and infrared proximity. Annotations consist of facing formulation, head and body orientation, and participant bounding boxes.

**Triadic Interaction** [61] is a 3D motion capture dataset containing 180 videos of a haggling game scenario with 120 participants. It captures triadic interactions in a lab setting, including video (500 views), depth (10 sensors), audio (23 mics), individual voice, and 3D point clouds. It provides benchmarks for predicting speaking status, social formation, and body gestures in multi-party interactions.

**ConfLab** The ConfLab [62] dataset is also established for F-formulation estimation similar to SALSA. The dataset comprises 1-hour overhead-view videos of people attending a conference. In addition to RGB videos, the dataset also collects audio signal, IMU data and Bluetooth proximity of via the wearable devices. The annotations include F-formulation, keypoints and speaking status of each participant.

**Multi-speaker Conversation** [63] contains egocentric video recordings of 50 participants divided into groups of 5, engaging in conversations around a table. Each video covers one group, and participants are instructed to listen or engage in their own group, allowing for the observation of attention labels. The dataset includes approximately 20 hours of egocentric videos with six-channel audio streams.

**VideoGazeSpeech** VideoGazeSpeech [64] is a recent dataset that investigates the correlation of speech and human's gaze in social scenarios. The dataset is built based on the data collected by Xu *et al.* [74], but VideoGazeSpeech only selects 29 videos with audio stream. Each video has 400 500 frames involving 2 4 people on average. Frame-level gaze target is provided on each frame.

**Werewolf Among Us (Exp.)** [65] is an expanded dataset of [72]. This dataset provides new benchmarks and labels for speaking target identification, pronoun coreference resolution, and mentioned player prediction. It enables a high level of understanding of multimodal conversation (*e.g.*, verbal, non-verbal cues) in multi-party settings. It provides labels for the 234 videos in the original Werewolf Among Us dataset.

### 3.3 Social Situation Analysis Datasets

There are social situation analysis datasets mainly for social relationship reasoning and social question answering. The performance of a model for social relation inference is commonly evaluated by classification accuracy. In terms of social question answering, a question is raised based on the social situation and a couple of answer candidates are provided. The AI models have to correctly parse the social scenes and select the correct answer(s) for the given question. The performance is usually measured by accuracy, i.e., the percentage of questions that are correctly answered. A variant for the question answering setting is generating answers in free-form language to the given questions instead of selecting from pre-defined answers. The performance is measured by BLEU [75], METEOR [76], ROUGE [77], BERTscore [78] and user study. Another dataset using werewolf games to evaluate the performance of models in learning the human's persuasion strategies and voting intentions [72]. They straightforwardly use accuracy as the metric for the two tasks.

**MovieGraphs** [66] contains 7637 movie clips annotated with graphs that capture who is present, their attributes, interactions, and relationships. Each graph consists of several node types including characters, their attributes, relationships, interactions, topics of interactions, reasons behind interactions, and time stamps. The dataset includes free-text labels and encompasses about 100 relationships, such as parent, spouse, and friend.

**TVQA** [67] is a dataset for video question answering that is built on 6 TV shows spanning 3 genres: crime shows, sitcoms and medical dramas. The dataset consists of 152.5K QA pairs that span 21,783 video clips from 925 episodes. Each video has 7 questions with 5 answer candidates, and only one is the correct answer. The length of the videos clips vary in between 60-90 seconds containing social interaction and activities.

**ViSR** The Video-based Social Relation dataset (ViSR) [68] contains over 8,000 video clips annotated with 8 social relation types derived from social domain theory. The dataset covers a variety of daily life social relations such as parent-offspring, couple, friends, colleagues etc. The videos are from movies and TV shows, providing diverse scenes and backgrounds.

**Social-IQ** [69] consists of a set of 7500 questions, 52500 answers (30000 correct and 22500 incorrect) covering a broad range of 1250 social videos. The videos in this dataset come from a diverse set of YouTube videos consisting of distinct characters across the videos. The questions and answers in this dataset are related to multiple modalities in social events. This dataset was further extended to Social-IQ 2.0 [79].

**LVU** [70] evaluates understanding of long-form content through 9 tasks on 30,000 video clips up to 3 minutes. Tasks include content analysis (*e.g.*, relationship, speaking style), user engagement prediction (*e.g.*, ratio, popularity), and movie metadata prediction (*e.g.*, genre, director). LVU contains 4 relationship types: 'friends', 'wife-and-husband', 'boyfriend-and-girlfriend', and 'ex-boyfriend-and-ex-girlfriend'.

**SMILE** The SMILE [71] dataset is a multimodal dataset for understanding laughter in videos with language models. The dataset comprises 887 video clips, each paired with a language description about the reason for laughter for the corresponding video clip. The video clips come from TED talks, where each video is between 10-90 seconds long, and sitcoms, where each video is 7-60 seconds long.

**Werewolf Among Us** [72] contains 252 videos and transcripts of social deduction games. One Night Ultimate Werewolf and The Resistance: Avalon games are obtained from YouTube and Ego4D [38]. Players form two teams and use persuasion strategies. Six typical persuasion strategies are pre-defined on the dataset. Annotations include persuasion strategies per utterance, player roles (start and end), and voting outcomes.



\end{document}